\documentclass[twocolumn,journal]{IEEEtran}
\usepackage[T1]{fontenc}
\usepackage[latin9]{inputenc}
\usepackage{color}
\usepackage{array}
\usepackage{pifont}
\usepackage{float}
\usepackage{textcomp}
\usepackage{mathrsfs}
\usepackage{amsmath}
\usepackage{amsthm}
\usepackage{amssymb}
\usepackage{stackrel}
\usepackage{graphicx}
\usepackage{setspace}
\usepackage[unicode=true,
 bookmarks=true,bookmarksnumbered=true,bookmarksopen=true,bookmarksopenlevel=1,
 breaklinks=false,pdfborder={0 0 0},pdfborderstyle={},backref=false,colorlinks=false]
 {hyperref}
\hypersetup{pdftitle={Your Title},
 pdfauthor={Your Name},
 pdfpagelayout=OneColumn, pdfnewwindow=true, pdfstartview=XYZ, plainpages=false}

\makeatletter

\providecommand{\tabularnewline}{\\}
\floatstyle{ruled}
\newfloat{algorithm}{tbp}{loa}
\providecommand{\algorithmname}{Algorithm}
\floatname{algorithm}{\protect\algorithmname}

\let\oldforeign@language\foreign@language
\DeclareRobustCommand{\foreign@language}[1]{%
  \lowercase{\oldforeign@language{#1}}}
\theoremstyle{plain}
\newtheorem{thm}{\protect\theoremname}
\theoremstyle{remark}
\newtheorem{rem}[thm]{\protect\remarkname}
\theoremstyle{plain}
\newtheorem{prop}[thm]{\protect\propositionname}

\usepackage[caption=false,font=footnotesize]{subfig}

\makeatother

\providecommand{\propositionname}{Proposition}
\providecommand{\remarkname}{Remark}
\providecommand{\theoremname}{Theorem}

\begin{document}
\title{Secrecy Rate of the Cooperative RSMA-Aided UAV Downlink Relying on
Optimal Relay Selection}
\author{\setstretch{0.98}\noindent Hamed~Bastami,~Majid~Moradikia,~Hamid~Behroozi,~\IEEEmembership{Member,~IEEE,}~and~Lajos~Hanzo,~\IEEEmembership{Life~Fellow,~IEEE}\thanks{Hamed~Bastami and Hamid~Behroozi are with the Department of Electrical
Engineering, Sharif University of Technology, Tehran, Iran, e-mail{\small{}s:
\{hamed.bastami@ee., behroozi@\}sharif.edu}}\thanks{Majid Moradikia is with Department of Data Science Worcester Polytechnic
Institute, Worcester, Massachusetts, e-mail: \{mmoradikia@wpi.edu\}}\thanks{Lajos Hanzo is with the University of Southampton, Southampton SO17
1BJ, U.K, e-mail: \protect\href{mailto:hanzo@soton.ac.uk}{hanzo@soton.ac.uk}.}}
\markboth{}{Hamed Bastami \MakeLowercase{\emph{et al.}}: Secrecy-Rate of the
Cooperative RSMA-Aided UAV Downlink Relying on Optimal Relay Selection}
\maketitle
\begin{abstract}
The Cooperative Rate-Splitting (CRS) scheme, proposed evolves from
conventional Rate Splitting (RS) and relies on forwarding a portion
of the RS message by the relaying users. In terms of secrecy enhancement,
it has been shown that CRS outperforms its non-cooperative counterpart
for a two-user Multiple Input Single Output (MISO) Broadcast Channel
(BC). Given the massive connectivity requirement of 6G, we have generalized
the existing secure two-user CRS framework to the multi-user framework,
where the highest-security users must be selected as the relay nodes.
This paper addresses the problem of maximizing the Worst-Case Secrecy
Rate (WCSR) in a UAV-aided downlink network where a multi-antenna
UAV Base-Station (UAV-BS) serves a group of users in the presence
of an external eavesdropper ($Eve$). We consider a practical scenario
in which only imperfect channel state information of $Eve$ is available
at the UAV-BS. Accordingly, we conceive a robust and secure resource
allocation algorithm, which maximizes the WCSR by jointly optimizing
both the Secure Relaying User Selection (SRUS) and the network parameter
allocation problem, including the RS transmit precoders, message splitting
variables, time slot sharing and power allocation. To circumvent the
resultant non-convexity owing to the discrete variables imposed by
SRUS, we propose a two-stage algorithm where the SRUS and network
parameter allocation are accomplished in two consecutive stages. With
regard to the SRUS, we study both centralized and distributed protocols.
On the other hand, for jointly optimizing the network parameter allocation
we resort to the Sequential Parametric Convex Approximation (SPCA)
algorithm. Our numerical results show that the proposed solution significantly
outperforms the existing benchmarks for a wide range of network loads
in terms of the WCSR. 
\end{abstract}


\section{Introduction}

\IEEEPARstart{6}{G} wireless communications are envisioned to support
the heterogeneous services of a massive number of connected users
with ultra-reliability, e.g., $1000\times$ higher mobile data volume
per geographical area, as well as $10\sim100\times$ more connected
devices, offered at efficient resources usage \cite{1,2}. These indicative
parameters are known as Key Performance Indicators (KPIs) of 6G. Upon
growing the dimension of the network and the number of connected users
over the limited shared spectrum, the problems of Inter-User Interference
(IUI) become exacerbated. To address these concerns, efficient Multiple
Access (MA) technologies, such as Rate-Splitting Multiple Access (RSMA),
have to be utilized \cite{3,4,5}. RSMA may be viewed as a generalized
generalized Non-Orthogonal Multiple Access (NOMA) and Space-Division
Multiple Access (SDMA) framework that is resilient against outdated
Channel State Information (CSI) \cite{4,5}. 
\begin{table*}[tbh]
\centering{}\caption{Overview of existing literature}
\begin{tabular}{|>{\raggedright}p{2.4cm}|>{\centering}m{1.3cm}|>{\centering}m{0.3cm}|>{\centering}m{0.3cm}|>{\centering}m{0.3cm}|>{\raggedright}m{0.3cm}|>{\centering}m{0.3cm}|>{\centering}m{0.3cm}|>{\raggedright}m{0.3cm}|>{\raggedright}m{0.3cm}|>{\raggedright}m{0.3cm}|>{\raggedright}m{0.3cm}|>{\raggedright}m{0.3cm}|>{\raggedright}m{0.3cm}|>{\raggedright}m{0.3cm}|>{\raggedright}m{0.3cm}|>{\raggedright}m{0.3cm}|>{\raggedright}m{0.3cm}|}
\hline 
{\small{}}%
\begin{tabular}{c}
{\small{}References$\Rightarrow$}\tabularnewline
\hline 
\hline 
{\small{}Keywords$\Downarrow$}\tabularnewline
\end{tabular} & {\small{}Proposed}\\
{\small{} Approach} & \raggedright{}{\small{}\cite{3}} & \raggedright{}{\small{}\cite{4}} & \raggedright{}{\small{}\cite{8}} & \raggedright{}{\small{}\cite{9}} & \raggedright{}{\small{}\cite{10}} & \raggedright{}{\small{}\cite{11}} & \raggedright{}{\small{}\cite{12}} & \raggedright{}{\small{}\cite{12.1}} & \raggedright{}{\small{}\cite{13.1}} & \raggedright{}{\small{}\cite{14}} & \raggedright{}{\small{}\cite{15}} & \raggedright{}{\small{}\cite{16}} & \raggedright{}{\small{}\cite{17}} & \raggedright{}{\small{}\cite{20}} & \raggedright{}{\small{}\cite{23}} & \raggedright{}{\small{}\cite{26}}\tabularnewline
\hline 
\hline 
\textbf{\small{}PLS} & $\checkmark$ &  &  & $\checkmark$ & $\checkmark$ & $\checkmark$ & $\checkmark$ & $\checkmark$ & $\checkmark$ & $\checkmark$ & $\checkmark$ & $\checkmark$ &  &  & $\checkmark$ &  & $\checkmark$\tabularnewline
\hline 
\textbf{\small{}Beamformer Design} & $\checkmark$ & $\checkmark$ & $\checkmark$ & $\checkmark$ & $\checkmark$ & $\checkmark$ &  & $\checkmark$ & $\checkmark$ & $\checkmark$ & $\checkmark$ & $\checkmark$ & $\checkmark$ & $\checkmark$ &  & $\checkmark$ & $\checkmark$\tabularnewline
\hline 
\textbf{\small{}Unknown $Eve$} &  &  &  &  &  & $\checkmark$ &  &  &  &  &  &  &  &  &  &  & \tabularnewline
\hline 
\textbf{\small{}FJ Design} & $\checkmark$ &  &  &  & $\checkmark$ & $\checkmark$ & $\checkmark$ & $\checkmark$ &  &  &  &  &  &  &  &  & \tabularnewline
\hline 
\textbf{\small{}SSRM} &  &  &  &  &  &  &  &  & $\checkmark$ &  & $\checkmark$ &  &  &  & $\checkmark$ &  & \tabularnewline
\hline 
\textbf{\small{}UAV-BS} & $\checkmark$ & $\checkmark$ &  &  &  &  & $\checkmark$ & $\checkmark$ & $\checkmark$ &  &  & $\checkmark$ &  &  & $\checkmark$ & $\checkmark$ & \tabularnewline
\hline 
\textbf{\small{}NOMA} &  & $\checkmark$ &  &  &  &  &  &  &  & $\checkmark$ &  &  & $\checkmark$ &  &  & $\checkmark$ & \tabularnewline
\hline 
\textbf{\small{}I-CSIT} & $\checkmark$ &  & $\checkmark$ &  &  &  &  & $\checkmark$ &  &  &  &  & $\checkmark$ &  &  & $\checkmark$ & \tabularnewline
\hline 
\textbf{\small{}SOPM} &  &  &  &  &  &  & $\checkmark$ &  &  & $\checkmark$ &  &  &  &  &  &  & \tabularnewline
\hline 
\textbf{\small{}RSMA} & $\checkmark$ &  & $\checkmark$ &  &  &  &  & $\checkmark$ & $\checkmark$ &  & $\checkmark$ & $\checkmark$ & $\checkmark$ & $\checkmark$ &  &  & $\checkmark$\tabularnewline
\hline 
\textbf{\small{}CRS} & $\checkmark$ &  &  &  &  &  &  & $\checkmark$ & $\checkmark$ &  & $\checkmark$ &  &  & $\checkmark$ &  &  & \tabularnewline
\hline 
\textbf{\small{}WCSRM} & $\checkmark$ &  & $\checkmark$ &  &  &  &  & $\checkmark$ &  &  &  & $\checkmark$ &  &  &  &  & $\checkmark$\tabularnewline
\hline 
\textbf{\small{}I-ECSIT} & $\checkmark$ &  &  &  &  &  &  & $\checkmark$ &  &  &  & $\checkmark$ &  &  &  &  & $\checkmark$\tabularnewline
\hline 
\textbf{\small{}SRUS} & $\checkmark$ &  &  &  &  &  &  &  &  &  &  &  &  &  &  &  & \tabularnewline
\hline 
\textbf{\small{}MU-CRS} & $\checkmark$ &  &  &  &  &  &  &  &  &  &  &  &  &  &  &  & \tabularnewline
\hline 
\end{tabular}
\end{table*}

Given the open nature of the wireless medium, Multi-User (MU) systems
are susceptible to especially security breaches, when a massive number
of connected users intend to utilize the same spectrum. Traditionally,
secure communication is achieved by employing cryptography encryption,
which-somewhat optimistically assumes limited computational capabilities
for the eavesdroppers ($Eves$), as well as perfectly secure key transfers
over the wireless medium. Yet, with the emergence of powerful quantum
computers, a persistent and sophisticated $Eve$ will be able to extract
important confidential information. To circumvent this concern, Physical
Layer (PHY)-Security (PLS), which relies on opportunistically exploiting
the random nature of the fading channels, has gained significant attention
\cite{6,7,8}. A pair of potent PLS designs rely on: 1) Optimizing
the active transmit precoder (beamformer) of a multi-antenna transmitter,
aimed at focusing the transmit power in the directions of legitimate
users, while minimizing the energy leakage to $Eves$ \cite{9,10};
2) intentionally broadcasting specifically designed Artificial Noise
(AN) everywhere except for focusing any potential $Eve$ \cite{11}.
However, both the performance and the design of the aforementioned
PLS schemes significantly depends on the accuracy of the knowledge
about the $Eve$\textquoteright s CSI at the legitimate transmitter
(E-CSIT) \cite{9,12}. 

To meet the demanding KPIs of 6G while securing the confidentiality
of the transmitted MUs\textquoteright{} messages, RSMA was recently
shown to be one of the most effective PLS techniques \cite{12}-\cite{15}.
By relying on a secure RSMA Transmit Precoder (TPC) a common message
which is constituted by a specific portion of the transmitted message,
is introduced at source with a twin-fold mission. More explicitly,
apart from serving as the desired message, it also acts as AN without
consuming extra power \cite{15}. This is in stark contrast to the
conventional AN design, where some portion of the transmit power budget
is allocated to AN, hence leading to inefficient usage of the available
power budget. Additionally, by optimizing the RSMA TPC at the Base-Station
(BS), we can deal with the effects of realistic imperfect channel
estimation at the Tx \cite{4,5,16}. As a beneficial extension of
RS, two-user Cooperative Rate-Splitting (CRS) has been proposed in
\cite{17} that outperforms its non-cooperative version \cite{16}
in terms of both its reliability \cite{18} and security \cite{12,14}.
Briefly, the CRS framework benefits from the cooperation of the legitimate
users, which are allowed to opportunistically forward their flawlessly
decoded common message to the distant user in two subsequent time-slots. 

Recently, Unmanned Aerial Vehicle (UAV)-aided communications have
attracted significant research interest \cite{20,21,22}, thanks to
their, on-demand coverage, and the availability of the Line of Sight
(LoS) links. In particular, the PLS attained may be readily improved
by UAVs upon detecting the $Eve$\textquoteright s location via the
UAV-mounted cameras or radar \cite{20}. In practice, only imperfect
E-CSIT may be available and thus the conventional PLS schemes no longer
perform at their best \cite{12,15}. Some of the associated challenges
have been addressed by the robust PLS solutions designed for UAV-enabled
scenarios in \cite{23}, \cite{24}. Recently, secure RS and CRS-aided
UAV networks relying on realistic imperfect CSIT have been studied
respectively in \cite{12,25}. However, the solutions proposed in
\cite{12} and \cite{25} are not applicable for multi-user networks,
where secure Relaying Users Selection (SRUS) protocols are necessitated
for security reasons. In other words, since all users may act potential
candidates for forwarding the common stream during the relaying phase,
the question arises: \textquotedbl{} \textit{How to beneficially select
the relaying users with the objective of enhancing the security?}\textquotedbl{}
\begin{table}[tbh]
\caption{\textcolor{black}{List of acronyms and abbreviations}}

\centering{}%
\begin{tabular}{l|r}
\hline 
\textbf{\small{}PLS} & {\small{}Physical Layer Security}\tabularnewline
\textbf{\small{}FJ} & {\small{}Friendly Jammer}\tabularnewline
\textbf{UAV-BS} & Unmanned Aerial Vehicle Base-Station\tabularnewline
\textbf{\small{}SIC} & {\small{}Successive Interference Cancellation}\tabularnewline
\textbf{\small{}I-ESIT} & {\small{}Imperfect $Eve$'s Channel State Information}\tabularnewline
\textbf{\small{}RSMA} & {\small{}Rate-Splitting Multiple Access}\tabularnewline
\textbf{\small{}MU-CRS} & {\small{}Multi User Cooperative Rate-Splitting}\tabularnewline
\textbf{\small{}NRS} & {\small{}Non-Cooperative Rate-Splitting}\tabularnewline
\textbf{\small{}NOMA} & {\small{}Non-Orthogonal Multiple Access}\tabularnewline
\textbf{\small{}SDMA} & {\small{}Space Division Multiple Access}\tabularnewline
\textbf{EIR} & Estimated Information Rates\tabularnewline
\textbf{\small{}WCSRM} & {\small{}Worst Case Secrecy Rate Maximization}\tabularnewline
\textbf{\small{}ACC} & Actual Channel Capacities\tabularnewline
\textbf{\small{}AN} & {\small{}Artificial Noise}\tabularnewline
\textbf{\small{}CEU} & Cell Edge Users \tabularnewline
\textbf{CCU} & Cell Center Users\tabularnewline
\textbf{WCSR} & Worst-Case Secrecy Rate\tabularnewline
\textbf{IUI} & Inter-User Interference\tabularnewline
\textbf{LoS} & Line of Sight\tabularnewline
\textbf{\small{}MISO-BC} & {\small{}Multi-Input Single-Output Broadcast Channel}\tabularnewline
\textbf{TPC} & Transmit Precoder\tabularnewline
\textbf{MA} & Multiple Access\tabularnewline
\textbf{SPCA} & Sequential Parametric Convex Approximation\tabularnewline
\textbf{\small{}SURS} & Secure Relaying Users Selection\tabularnewline
\hline 
\end{tabular}
\end{table}
 None of the above secure UAV-RSMA designs have addressed this important
research question. Hence our objective is to close this knowledge
gap. Furthermore, to circumvent the deleterious impact of imperfect
CSIT, in contrast to \cite{19}, we have conceived the worst-case
robust designs as well. Explicitly, this is the first work that investigates
the robust and secure design of the generalized multi-user C-RSMA
downlink of UAV networks. To gain deeper insights, the novelty of
the proposed approach is boldly and explicitly contrasted to the state-of-the-art
in Table I at a glance. Table II provides a list of acronyms used
in this paper.

Against this background, the detailed contributions of our work are
summarized as follows:
\begin{figure}[tbh]
\centering{}\includegraphics[viewport=340bp 50bp 760bp 440bp,clip,scale=0.62]{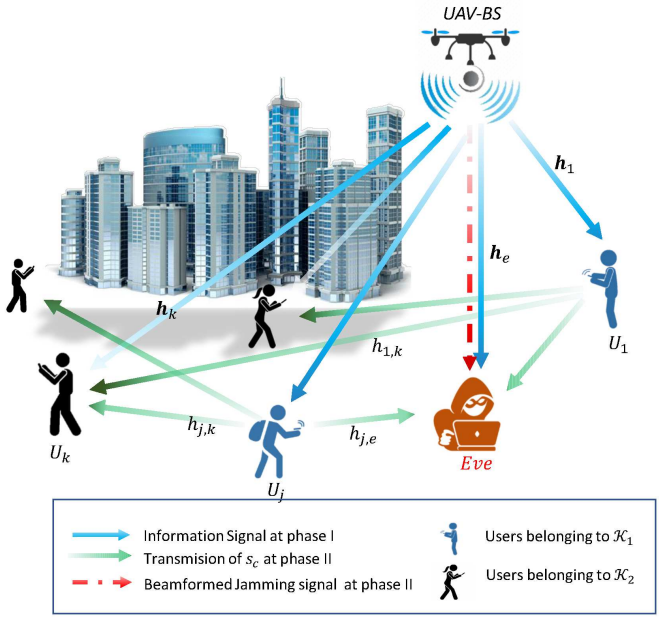}\caption{\label{fig:The-proposed-secure}The proposed cooperative rate-splitting
UAV network.}
\end{figure}

\begin{itemize}
\item By relying on the two-stage multi-user C-RSMA philosophy, we follow
the secrecy policy of \cite{12} for safeguarding the first cooperative
phase through the secure design of the common RSMA TPC. Interestingly,
by decoding and forwarding the common stream, two opportunities are
provided to take full advantage of the common stream, hence further
enhancing the secrecy. However, for preventing the interception of
the common message during the second phase, the AN is directed towards
the $Eve$ by designing a robust Maximum Ratio Transmission (MRT)
beamformer, which is achieved by harnessing an idle UAV to serve as
a jammer during the relaying phase. 
\item To safeguard confidentiality in the presence of $Eve$, we conceive
a generalized multi-user C-RSMA secrecy design, where we optimize
the TPCs, the message split, the time slot allocation, the jamming
power as well as SRUS  to guarantee both secrecy fairness among all
users and robustness against imperfect E-CSIT. To satisfy our secrecy
fairness objective, we formulate the associated Worst-Case Secrecy
Rate (WCSR) maximization problem optimized subject to specific secrecy
policy constraints due to the common message rate as well as the power
budget constraints imposed on the UAV-BSs during both the transmission
and phases.
\item To circumvent the discrete nature of the SRUS process as well as the
non-convexity of our problem, we propose a two-stage algorithm where
SRUS and network parameter optimization is accomplished in two consecutive
stages. As for the SRUS, we study both the centralized and distributed
protocols. On the other hand, for jointly optimizing the precoders,
message split, time slot allocation, and jamming power we resort to
the Sequential Parametric Convex Approximation (SPCA) algorithm. This
allows us to strike a trade-off between accuracy of approaching the
optimal solution and the computational complexity.
\item Numerical results show that by applying the proposed solution, the
WCSR is substantially boosted over that of the non-cooperative benchmarks
over a wide range of network loads.
\end{itemize}
The rest of this paper is organized as follows. The system model,
signal representation, corresponding achievable information rate and
other preliminaries are provided in Section \ref{sec:II}. Section
\ref{sec:III} formulates our robust WCSR maximization problem and
optimal SRUS protocol. The proposed SPCA-based solution, the associated
feasible initialization procedures, and our complexity as well as
convergence analysis are provided in Section \ref{sec:IV}. In Section
\ref{sec:V}, our simulation results are presented and conclude in
Section \ref{sec:VI}. Finally, the Appendices and proofs of the lemmas
are provided. 

\textit{Notation:} Vectors and matrices are denoted by lower-case
and upper-case boldface symbols, respectively; $\left(.\right)^{\mathrm{T}}$,
$\left(.\right)^{\mathrm{*}}$, $\left(.\right)^{\mathrm{H}}$, and
$\left(.\right)^{\mathrm{-1}}$ denote the transpose, conjugate, conjugate
transpose, and inverse of a matrix respectively; $\mathfrak{R}e(.)$
denote the real part of a complex variable, and $\mathfrak{I}m(.)$
denote the imaginary part of a complex variable; We use $\mathbb{E}{\left\{ \cdot\right\} }$
and $\triangleq$ to denote the expectation and definition operations,
respectively; A complex Gaussian random variable with mean $\mathit{\mu}$
and variance $\sigma^{2}$ reads as $\mathcal{C}\mathcal{N}\left(\mathit{\mu},\sigma^{2}\right)$;
The notation $\mathbf{I}_{N}$ denotes the $N\times N$ identity matrix;
$\mathbb{R}^{N\text{\texttimes}1}$ and $\mathbb{C}^{N\times1}$ denote
the set of $N$-dimensional standard real and complex Gaussian random
variable, respectively; $\mathbb{C}^{N\times N}$ stands for an $N\times N$
element standard complex Gaussian random matrix whose real and imaginary
parts are independent normally distributed random variables with mean
zero and variance $\frac{1}{2}$; Finally, the entry in the $i$-th
row and $j$-th column of a matrix $\mathbf{H}$ is represented by
$\mathbf{H}\left[i,j\right]$. 

\section{System Model and Preliminaries \label{sec:II}}

We commence by introducing the principles of C-RSMA and the channel
models. The system model considered is illustrated in Fig. \ref{fig:The-proposed-secure},
in which a multi-antenna UAV-BS aims for concurrently serving $K$
legitimate users on the ground indexed by the set $\mathcal{K}={\left\{ 1,2,...,K\right\} }$
in the presence of an $Eve$, who silently engages in covert wiretapping.
While the UAV-BS is equipped with $N_{t}$ transmit antennas, the
terrestrial nodes are assumed to be single-antenna devices. For simultaneously
serving multiple users, the UAV-BS is supported by RSMA. Note that,
as it will be discussed later in detail, to take full advantage of
RSMA, we exploit the more sophisticated C-RSMA technique in this paper,
in which two groups of users are considered for each transmission
slot. Thus, the users are divided into two different groups: Cell
Center Users (CCU) indexed by the set $\mathcal{K}_{1}$, and Cell
Edge Users (CEU) indexed by the set $\mathcal{K}_{2}$, so that we
have $\mathcal{K}_{1}\bigcup\mathcal{K}_{2}=\mathcal{K}$. We assume
furthermore that the channel conditions of the users of $\mathcal{K}_{1}$
are superior to the channels of those belonging to $\mathcal{K}_{2}$.

\subsection{Principles of C-RSMA}

Based on the principle of C-RSMA, the signal transmission is completed
in two consecutive phases: 
\begin{figure}[tbh]
\centering{}\includegraphics[viewport=70bp 420bp 350bp 724bp,clip,scale=0.9]{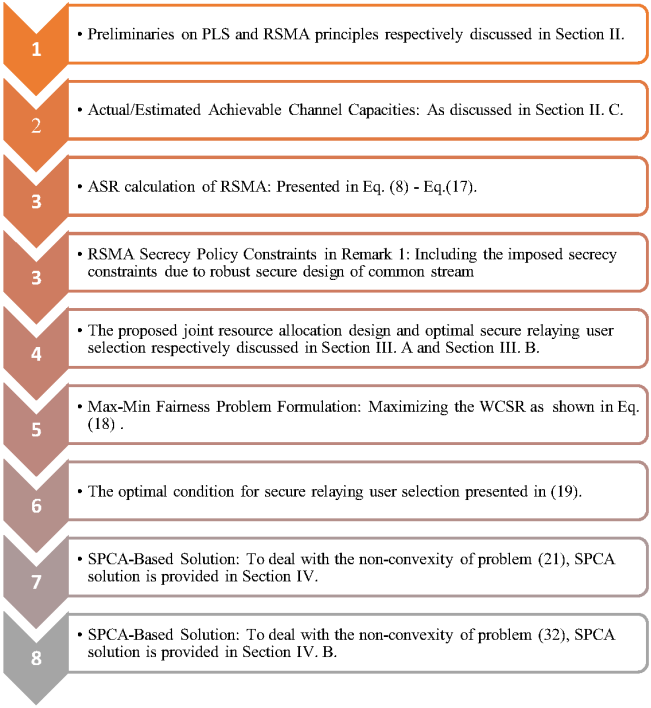}\caption{\textcolor{black}{\label{fig:Flow-of-the}Flow of the mathematical
analysis}}
\end{figure}

\begin{enumerate}
\item Broadcasting Phase (BP), where the UAV-BS transmits the RSMA precoded
signal towards the terrestrial nodes through the typical Multiple
Input Single Output (MISO) Downlink (DL) channels, i.e., represented
by $\textrm{UAV-BS}\rightarrow\left\{ \left.U_{k}\right|_{k\in\mathcal{K}},\ Eve\right\} $, 
\item Relaying Phase (RP), where users within $\mathcal{K}_{1}$, cooperatively
forward the signals to CEUs, i.e., $\left\{ \left.U_{k}\right|_{k\in\mathcal{K}_{1}}\right\} \rightarrow\left\{ \left.U_{k}\right|_{k\in\mathcal{K}_{2}},\ Eve\right\} $. 
\end{enumerate}
The number of time-slots allocated to the two phases may not be equal.
Hence, we introduce a dynamic time-slot allocation parameter $0<\theta\le1$,
where the fraction $\theta$ of time is dedicated for the BP and the
remaining portion $\left(1-\theta\right)$ is allocated for the RP. 

Following the 1-layer RSMA principle \cite{18}, the confidential
messages of each legitimate user $\left.\mathcal{W}_{k}\right|_{k=1}^{K}$
are split into the common parts $\left.\mathcal{W}_{k}^{c}\right|_{k=1}^{K}$
and the private parts $\left.\mathcal{W}_{k}^{p}\right|_{k=1}^{K}$.
The so-obtained common parts are then incorporated into a common codebook
for generating the unified common stream $s_{c}$. By contrast, $\left.\mathcal{W}_{k}^{p}\right|_{k=1}^{K}$
are independently encoded into the private streams $\left.s_{k}\right|_{k=1}^{K}$,
respectively. If we define $\mathbf{s}\triangleq\left[s_{1},\ s_{2},\ \ldots,s_{K}\right]^{T}$
with a normalized power of $\mathbb{E}\left\{ \mathbf{s}\mathbf{s}^{H}\right\} =\mathbf{I}_{K+1}$
and $\mathbf{P}\triangleq\left[\mathbf{p}_{{c}},\mathbf{p}_{\mathbf{1}},\ldots,\mathbf{p}_{K}\right]^{T}$
as the TPC matrix adopted by the UAV-BS, where $\mathbf{p}_{c}$ and
$\left.\mathbf{p}_{k}\right|_{k=1}^{K}$ respectively stand for the
parts corresponding to the common and private streams, then the RSMA
DL signal transmitted during the BP is given by:
\begin{equation}
\mathbf{x}^{\left(1\right)}=\mathbf{P}\mathbf{s}=\stackrel[k\in\mathcal{K}]{}{\sum}\mathbf{p}_{k}s_{k}+\mathbf{p}_{c}s_{c}.\label{eq:1}
\end{equation}

At the receiver side, each user first recovers $\mathcal{W}_{k}^{c}$
from the detected $\mathbf{s}_{c}$ and then removes the common stream
by performing Successive Interference Cancellation (SIC). Then, each
user detects its corresponding private stream. By combining $\mathcal{W}_{k}^{c}$
with the respective private message, each user can retrieve its original
intended message $\mathcal{W}_{k}$. During the RP, only the decoded
common streams are forwarded to the CEUs. According to the above discussion,
two opportunities are provided for $Eve$ to infer the information
during the two phases of transmissions. To guarantee the security,
while each user is expected to be able to decode $s_{c}$, as a part
of its original message, the corresponding common TPC, i.e., $\mathbf{p}_{c}$,
must be designed for ensuring that $s_{c}$ simultaneously plays the
role of undecodable jamming signal at $Eve$. Since the UAV-BS is
idle during the RP, we assume that it serves as a friendly jammer
during this phase to further enhance the secrecy. 

\subsection{Channel Models }

We focus on a quasi-static fading environment and denote the channel
coefficients of the links spanning from the UAV-BS to terrestrial
legitimate nodes and $Eve$ respectively by $\mathbf{h}_{k}\in\mathbb{C}^{N_{t}\times1}$
and $\mathbf{h}_{e}\in\mathbb{C}^{N_{t}\times1}$. These channels
are modeled as $\mathbf{h}_{n}\triangleq PL\left(d_{n}\right)\,\mathbf{n}_{n},\ \forall n\in\left\{ \mathcal{K},e\right\} $,
where $PL\left(d_{n}\right)$ represent the path-loss characterized
by the Aerial-to-Ground (A2G) distance $d_{n}$, and $\mathbf{n}_{n}~\mathcal{CN}\left(0,\ \mathbf{I}_{N_{t}}\right)$
represents the corresponding small scale fading. Similarly, $h_{j,i}\triangleq PL\left(d_{j,i}\right)\,n_{j,i},\ \forall j\in\left\{ \mathcal{K}_{1}\right\} ,\forall i\in\left\{ \mathcal{K}_{2},e\right\} $
stands for the Single Input Single Output (SISO) channel from CCUs
and CEUs as well as $Eve$ during RP. 

Since large-scale fading varies smoothly, we suppose that the UAV-BS
can obtain the path-loss variable perfectly for the entire links.
Additionally, for the A2G legitimate links $\textrm{UAV}\rightarrow\left\{ \left.U_{k}\right|_{k\in\mathcal{K}}\right\} $,
we assume that the small-scale fading component can be obtained accurately
by frequently sending handshaking signals. However, concerning the
$Eve$, as an untrustworthy subscriber who does not regularly interact
with the UAV-BS, its CSI would be outdated at the UAV-BS owing to
signaling delays, i.e., Imperfect E-CSIT, although $Eve$ might be
able to perfectly estimate its corresponding channels in the worst-case
secrecy scenario. Thus, for the illegitimate links during both phases
$\left\{ \textrm{UAV},\left.U_{k}\right|_{k\in\mathcal{K}_{1}}\right\} \rightarrow Eve$
we assume that the large-scale fading can be estimated perfectly,
and imperfection can only contaminate the small-scale fading component.
To characterize the imperfect E-CSIT, we utilize the worst-case model
of \cite{26,27}, by which the small-scale fading coefficients of
the $\textrm{UAV}\rightarrow Eve$ and $\left\{ \left.U_{k}\right|_{k\in\mathcal{K}_{1}}\right\} \rightarrow Eve$
channels are formulated as follows: 
\begin{align}
\mathbf{n}_{e} & \triangleq\hat{\mathbf{n}}_{e}+\mathbf{\Delta n}_{e},\,\,\varTheta_{e}=\left\{ \mathbf{\Delta n}_{e}\in\mathbb{C}^{N_{t}\times1}:\,\left\Vert \mathbf{\Delta n}_{e}\right\Vert ^{2}\leq N_{t}\zeta^{2}\right\} \text{,}\label{eq:2}\\
n_{j,e} & \triangleq\hat{n}_{j,e}+\triangle n_{j,e},\,\,,\varTheta_{h_{j,e}}=\left\{ \triangle n_{j,e}:\,\left|\triangle n_{j,e}\right|^{2}\leq\zeta^{2}\right\} ,\nonumber 
\end{align}
$\,\forall\,j\in\mathcal{K}_{1}$, where $\hat{\mathbf{n}}_{e}$ and
$\hat{n}_{j,e}$ are respectively the estimated small-scale fading
of $Eve$ available for the UAV-BS. Furthermore, $\left.U_{j}\right|_{j\in\mathcal{K}_{1}}$,
and $\mathbf{\Delta n}_{e}$ and $\triangle n_{j,e}$ respectively
stand for the unknown channel uncertainty corresponding to $\mathbf{n}_{e}$
and $n_{j,e}$. Still referring to \eqref{eq:2}, $\zeta$ specifies
the radius of the bounded error regions of $\triangle n_{j,e}$, whilst
for $\mathbf{\Delta n}_{e}$ the error is a region bounded with radius
$\sqrt{N_{t}}\zeta$. Specifically, $\zeta>0$ denotes the size of
the uncertainty region of the small-scale fading estimate of the $Eve$.
Therefore, the estimated gains of the $\textrm{UAV}\rightarrow Eve$
and $\left.U_{j}\right|_{j\in\mathcal{K}_{1}}\rightarrow Eve$ channel
are ${\hat{\mathbf{h}}_{e}}=\mathrm{PL}\left(d_{e}\right)\,\hat{\mathbf{n}}_{e}$
and ${\hat{h}_{j,e}}=\mathrm{PL}\left(d_{j,e}\right)\,\hat{n}_{j,e}$. 

\subsection{Performance Metrics and Constraints}

In this section, we derive the Achievable Secrecy Rate (ASR) as the
objective function considered followed by the constraints imposed,
which must be taken into account in our design. Before proceeding,
we have provided a flow-diagram in Fig. \ref{fig:Flow-of-the} to
show the flow of the analysis described in the sequel. Again, the
worst-case secrecy scenario is considered in this paper, where only
imperfect E-CSIT is available at the UAV-BS. Given this perspective,
while $Eve$ can achieve the actual channel capacity, the UAV-BS is
only capable of achieving the estimated information rate. In the following,
we derive the Actual Channel Capacities (ACC) as well as the Estimated
Information Rates (EIR). 

\subsubsection{Received Signal Models}

During the BP, the UAV-BS broadcasts the RSMA signal $\mathbf{x}^{\left(1\right)}$
and the terrestrial nodes, i.e., $\left\{ \left.U_{k}\right|_{k\in\mathcal{K}}\right\} $
and $Eve$, will respectively receive the following signals:
\begin{equation}
{y_{k}}^{(1)}={\mathbf{h}_{k}}^{H}\mathbf{x}^{\left(1\right)}+{z_{k}}^{(1)},\,\forall\,k\in\mathcal{K}\label{eq:3}
\end{equation}
\begin{equation}
{y_{e}}^{(1)}={\mathbf{h}_{e}}^{H}\mathbf{x}^{\left(1\right)}+{z_{e}}^{(1)},\label{eq:4}
\end{equation}
where $z_{k}^{(1)}~\mathcal{CN}\left(0,\sigma_{k}^{2}\right)$ and
$z_{e}^{(1)}~\mathcal{CN}\left(0,\sigma_{e}^{2}\right)$ respectively
stand for the Additive White Gaussian Noise (AWGN). During the RP,
the $j$-th CCU re-encodes the $s_{c}$ and forwards it towards the
CEUs at a power of $p_{j}$. Concurrently, using the estimated channel
${\hat{\mathbf{h}}}_{e}$, the UAV-BS assigns the beamforming vector
$\mathbf{\hat{p}}_{z}\triangleq\frac{\mathbf{\hat{h}}_{e}^{H}}{\left\Vert \mathbf{\hat{h}}_{e}\right\Vert }$
directed towards the $Eve$ with the power of $p_{z}$ to cover the
transmission of $s_{c}$ against $Eve$. Hence, the signals transmitted
during the RP respectively from the UAV-BS and the $j$-th CCU are
given by: 
\begin{gather}
{x_{z}}^{(2)}=\sqrt{p_{z}}\mathbf{\hat{p}}_{z}z,\,\mathrm{\textrm{and}}\,{x_{j}}^{(2)}=\sqrt{p_{j}}\,s_{c}\,\forall\,j\in\mathcal{K}_{1}.\label{eq:5}
\end{gather}

Subsequently, the signals received by $k$-th CEU and $Eve$ in the
RP become:
\begin{gather}
{y_{k}}^{(2)}=\underset{j\in\mathcal{K}_{1}}{\sum}{h_{j,k}}^{H}{x_{j}}^{(2)}+\sqrt{p_{z}}{\mathbf{h}_{k}^{H}}\mathbf{\hat{p}}_{z}z+{z_{k}}^{(2)},\,\,\forall\,k\in\mathcal{K}_{2}\label{eq:6}
\end{gather}
\begin{gather}
{y_{e}}^{(2)}=\underset{j\in\mathcal{K}_{1}}{\sum}{h_{j,e}}^{H}{x_{j}}^{(2)}+\sqrt{p_{z}}{\mathbf{h}_{e}^{H}}\mathbf{\hat{p}}_{z}z+{z_{e}}^{(2)},\label{eq:7}
\end{gather}
where ${z_{k}}^{(2)}\sim\mathcal{CN}(0,\sigma_{k}^{2})$ and ${z_{e}}^{(2)}\sim\mathcal{CN}(0,\sigma_{e}^{2})$
are the AWGN at the legitimate user $\left\{ \left.U_{k}\right|_{k\in\mathcal{K}_{2}}\right\} $
and $Eve$, respectively. 

\subsubsection{ACC and EIR Analysis}

The CCUs first decode $s_{c}$, while treating the signals corresponding
to all the private messages as noise. Accordingly, the channel capacities
achieved by $\left\{ \left.U_{k}\right|_{k\in\mathcal{K}}\right\} $
and $Eve$ in decoding $s_{c}$ during the BP is given by:
\begin{gather}
{R_{c,k}}^{(1)}=\theta\log_{2}\left(1+\frac{\left|{\mathbf{h}_{k}}^{H}\mathbf{p}_{c}\right|^{2}}{\underset{j\in\mathcal{K}}{\sum}\left|{\mathbf{h}_{k}}^{H}\mathbf{p}_{j}\right|^{2}+\sigma_{k}^{2}}\right),\label{eq:8}
\end{gather}
\begin{gather}
{R_{c,e}}^{(1)}=\theta\log_{2}\left(1+\frac{\left|{\mathbf{h}_{e}}^{H}\mathbf{p}_{c}\right|^{2}}{\underset{j\in\mathcal{K}}{\sum}\left|{\mathbf{h}_{e}}^{H}\mathbf{p}_{j}\right|^{2}+\sigma_{e}^{2}}\right),\label{eq:9}
\end{gather}
$\forall\,k\in\mathcal{K}$. Under the assumption of perfect SIC,
the decoded common message is fully eliminated from the received signal
and then each user decodes the intended private message by treating
the remaining interference inflicted by the other private messages
as a noise. Therefore, only the unintended private stream can be considered
as interference and the ACC of decoding the private message by $\left\{ \left.U_{k}\right|_{k\in\mathcal{K}}\right\} $
is calculated as:
\begin{gather}
{R_{p,k}}=\theta\log_{2}\left(1+\frac{\left|{\mathbf{h}_{k}}^{H}\mathbf{p}_{k}\right|^{2}}{\underset{j\in\mathcal{K},j\neq k}{\sum}\left|{\mathbf{h}_{k}}^{H}\mathbf{p}_{j}\right|^{2}+\sigma_{k}^{2}}\right),\label{eq:10}
\end{gather}
$\,\forall\,k\in\mathcal{K}$. Upon using the RSMA technique, an appropriate
secrecy policy is to design the common TPC for ensuring that $Eve$
is unable to decode $s_{c}$. By doing so, $s_{c}$ can no longer
be eliminated through the preceding SIC block and its corresponding
term is considered as interference at $Eve$. Thus, considering that
$Eve$ has a single chance of deciphering the private messages during
the first phase, the SINR associated with the detection of the private
stream of $\left.U_{k}\right|_{k\in\mathcal{K}}$, while treating
the private steam of the other users $\left.U_{j}\right|_{j\in\mathcal{K},\ j\neq k}$
as well as the common stream as interference, may be expressed as:
\begin{gather}
{R_{k,e}}=\theta\log_{2}\left(1+\frac{\left|{\mathbf{h}_{e}}^{H}\mathbf{p}_{k}\right|^{2}}{\left|{\mathbf{h}_{e}}^{H}\mathbf{p}_{c}\right|^{2}+\underset{j\in\mathcal{K},j\neq k}{\sum}\left|{\mathbf{h}_{e}}^{H}\mathbf{p}_{j}\right|^{2}+\sigma_{e}^{2}}\right).\label{eq:11}
\end{gather}

During the RP, the ACC in decoding the $s_{c}$ at $\left.U_{k}\right|_{k\in\mathcal{K}_{2}}$
and $Eve$ are respectively given by:
\begin{gather}
{R_{c,k}}^{(2)}=\left(1-\theta\right)\log_{2}\left(1+\frac{\underset{j\in\mathcal{K}_{1}}{\sum}p_{j}\left|{h_{j,k}}\right|^{2}}{p_{z}\left\Vert {\mathbf{h}_{k}}^{H}\mathbf{\hat{p}}_{z}\right\Vert ^{2}+\sigma_{k}^{2}}\right),\label{eq:12}
\end{gather}
\begin{gather}
{R_{c,e}}^{(2)}=\left(1-\theta\right)\log_{2}\left(1+\frac{\underset{j\in\mathcal{K}_{1}}{\sum}p_{j}\left|{h_{j,e}}\right|^{2}}{p_{z}\left\Vert {\mathbf{h}_{e}}^{H}\mathbf{\hat{p}}_{z}\right\Vert ^{2}+\sigma_{e}^{2}}\right).\label{eq:13}
\end{gather}

Notably, the users in $\mathcal{K}_{2}$ and $Eve$ combine the decoded
common message in both phases. However, while the achievable capacity
of decoding the common message at all $\left.U_{k}\right|_{k\in\mathcal{K}}$
is limited by the worst-case user, i.e., by user receiving at the
minimum SINR in detecting $s_{c}$ during both phases, $Eve$ tries
to infer $s_{c}$ up to the sum-capacity of both phases. Accordingly,
the achievable rate of decoding $s_{c}$ at $\left.U_{k}\right|_{k\in\mathcal{K}}$
and $Eve$ are given as:
\begin{align}
{R_{c}} & =\min\left\{ \underset{{R_{c,\mathcal{K}_{1}}}}{\underbrace{\left.{R_{c,k}}^{(1)}\right|_{\forall\,k\in\mathcal{K}_{1}}}},\underset{{R_{c,\mathcal{K}_{2}}}}{\underbrace{\left.\left({R_{c,k}}^{(1)}+{R_{c,k}}^{(2)}\right)\right|_{\forall\,k\in\mathcal{K}_{2}}}}\right\} \nonumber \\
\, & =\min\left\{ {R_{c,\mathcal{K}_{1}}},{R_{c,\mathcal{K}_{2}}}\right\} .\label{eq:14}
\end{align}

Then, due to the deleterious impact of imperfect E-CSIT, the EIR corresponding
to the $Eve$\textquoteright s link from the UAV-BS viewpoint is formulated
as follows:
\begin{equation}
{\hat{R}_{c,e}}{}^{(1)}=\theta\log_{2}\left(1+\frac{\left|{\hat{\mathbf{h}}_{e}}^{H}\mathbf{p}_{c}\right|^{2}}{\underset{j\in\mathcal{K}}{\sum}\left|{\hat{\mathbf{h}}_{e}}^{H}\mathbf{p}_{j}\right|^{2}+\sigma_{e}^{2}}\right),\label{eq:15}
\end{equation}
\begin{equation}
{\hat{R}_{c,e}}{}^{(2)}=\left(1-\theta\right)\log_{2}\left(1+\frac{\underset{j\in\mathcal{K}_{1}}{\sum}p_{j}\left|{\hat{h}_{j,e}}\right|^{2}}{p_{z}\left\Vert {\hat{\mathbf{h}}_{e}}^{H}\mathbf{\hat{p}}_{z}\right\Vert ^{2}+\sigma_{e}^{2}}\right),\label{eq:16}
\end{equation}
\begin{equation}
{\hat{R}_{k,e}}=\theta\log_{2}\left(1+\frac{\left|{\hat{\mathbf{h}}_{e}}^{H}\mathbf{p}_{k}\right|^{2}}{\left|{\hat{\mathbf{h}}_{e}}^{H}\mathbf{p}_{c}\right|^{2}+\underset{j\in\mathcal{K},j\neq k}{\sum}\left|{\hat{\mathbf{h}}_{e}}^{H}\mathbf{p}_{j}\right|^{2}+\sigma_{e}^{2}}\right).\label{eq:17}
\end{equation}

\begin{rem}[Secrecy Policy]
\label{rem1} To ensure that the common message is decodable for
each legitimate user, the actual transmission rate of the common message
$r_{c}$ should satisfy the condition $r_{c}\ \le\ R_{c}$. On the
other hand, we aim for designing $\mathbf{p}_{c}$ at the UAV-BS,
so that the common message cannot be decoded and plays the role of
interference at the $Eve$. For this purpose, we have to satisfy another
conditionwe have to satisfy another condition, namely $r_{c}>\ R_{c,e}$.
As a result, the corresponding SINR of the private message at the
$Eve$ is degraded by the interference caused by the undecodable common
message.
\end{rem}
~
\begin{rem}[Common Message Considerations]
 $R_{c}$ is shared among all users to see whether the condition
$C_{1}$, pointed out in Remark \ref{rem1}, is satisfied or not.
Thus, the rate contributions of each user in transmitting $\mathcal{W}_{k}^{c}$,
denoted by $\mathcal{C}_{k}^{c}$, should be tuned at the UAV-BS so
that we have $R_{c}=\sum_{k\in\mathcal{K}}\mathcal{C}_{k}^{c}$, where
$\mathcal{C}_{k}^{c}=x_{k}R_{c}$. The weighting factors $x_{k}\in\left(0,1\right)$
associated with $\sum_{k\in\mathcal{K}}x_{k}=\ 1$ enable us to flexibly
adjust the significance of each user as part of the common secrecy
rate enhancement. 
\end{rem}
The total achievable secrecy rate of $k$-th user is given by $\mathcal{R}_{k}^{\text{sec}}=\mathcal{R}_{k,c}^{\text{sec}}+\mathcal{R}_{k,p}^{\text{sec}}$
, where $\mathcal{R}_{k,c}^{\text{sec}}=x_{k}\left[R_{c}-R_{c,e}\right]^{+}$
and $\mathcal{R}_{k,p}^{\text{sec}}=\left[R_{p,k}-R_{k,e}\right]^{+}$
represent the achievable secrecy rate of the common message and private
message transmitted to the $k$-th user, respectively.

\section{Proposed Joint Resource Allocation and SRUS \label{sec:III}}

Under the realistic imperfect CSIT assumption, the performance of
the MRT beamformer ${\hat{\mathbf{p}}}_{z}$ is degraded. To mitigate
this deleterious impact and to guarantee a robust design, we will
aim for maximizing the minimum ASR over all possible CSI uncertainties.
On the other hand, the system's objective is to maximize the minimum
ASR among all legitimate users, with the objective of maintaining
secrecy fairness. 

\subsection{Optimization Problem }

Based on the above discussion, the proposed joint resource allocation
design comprised of jointly optimizing the TPC, the common message
split, the time slot allocation, and SRUS  with the objective of maximizing
the minimum ASR among all users subject to a transmit power constraint
at the UAV-BS as well as the RSMA secrecy constraints is formulated
by: 
\begin{equation}
\underset{\mathbf{P},\theta,\mathbf{\boldsymbol{\chi}},\mathcal{K}_{1},\left\{ p_{j}\right\} _{\forall j\in\mathcal{K}_{1}},p_{z}}{\max}\left(\underset{k\in\mathcal{K}}{\min}\left(\begin{array}{cc}
\underset{\mathbf{\Delta n}_{e},\left\{ \triangle n_{j,e}\right\} _{\forall j\in\mathcal{K}_{1}}}{\min} & \left\{ \mathcal{R}_{k}^{\textrm{sec}}\right\} \end{array}\right)\right)\,\,\label{eq:18}
\end{equation}
s.t.$\textit{ }$

\begin{tabular}{>{\raggedright}p{15cm}}
$C_{1}:$ $r_{c}\leq R_{c}$,\vspace{0.1cm}
\tabularnewline
\end{tabular}

\begin{tabular}{>{\raggedright}p{15cm}}
$C_{2}:$ $R_{c}=\min\left\{ {R_{c,\mathcal{K}_{1}}},{R_{c,\mathcal{K}_{2}}}\right\} $,\vspace{0.1cm}
\tabularnewline
\end{tabular}

\begin{tabular}{>{\raggedright}p{15cm}}
$C_{3}:$ $r_{c}\geq\underset{\mathbf{\Delta n}_{e},\left\{ \triangle n_{j,e}\right\} _{\forall j\in\mathcal{K}_{1}}}{\max}{R_{c,e}}$,\vspace{0.1cm}
\tabularnewline
\end{tabular}

\begin{tabular}{>{\raggedright}p{15cm}}
$C_{4}:$ $\textrm{Tr}(\mathbf{P}{\mathbf{P}}^{H})\le P_{t},$\vspace{0.1cm}
\tabularnewline
\end{tabular}

\begin{tabular}{>{\raggedright}p{15cm}}
$C_{5}:$ $0\le p_{z}\le\bar{P}_{z}$\vspace{0.1cm}
\tabularnewline
\end{tabular}

\begin{tabular}{>{\raggedright}p{15cm}}
$C_{6}:$ $0\le p_{j}\le\bar{P}_{J}\,\,\forall j\in\mathcal{K}_{1}$\vspace{0.1cm}
\tabularnewline
\end{tabular}

\begin{tabular}{>{\raggedright}p{15cm}}
$C_{7}:$ $C_{k}\geq0,\,\,\forall k\in\mathcal{K}$\vspace{0.1cm}
\tabularnewline
\end{tabular}

\begin{tabular}{>{\raggedright}p{15cm}}
$C_{8}:$ $\mathcal{K}_{1}\subset\mathcal{K},\,\mathcal{K}_{2}=\mathcal{K}\setminus\mathcal{K}_{1}$\vspace{0.1cm}
\tabularnewline
\end{tabular}

\begin{tabular}{>{\raggedright}p{15cm}}
$C_{9}:$ $0<\theta\leq1,$\vspace{0.1cm}
\tabularnewline
\end{tabular}

\noindent where ${\boldsymbol{\chi}}\triangleq\left[x_{1},\ldots,x_{K}\right]$,
and $P_{t}$ stands for the transmit power budget at the UAV-BS. Note
that, once $\mathcal{K}_{1}$ is determined by solving the optimization
problem, $\mathcal{K}_{2}\triangleq{\mathcal{K}}\backslash{\mathcal{K}}_{1}$
also becomes specified, and thus only $\mathcal{K}_{1}$ is considered
as an optimization variable. The problem formulated is a mixed integer
non-convex problem due to the discontinuous nature of the variable
$\mathcal{K}_{1}$. The optimal SRUS problem itself imposes high computational
complexity and the resultant cost increases upon growing the number
of users. To circumvent this difficulty, we propose a low-cost algorithm
including the following two main steps. Firstly, SRUS protocol is
performed to find optimum $\mathcal{K}_{1}^{*}$ and $\mathcal{K}_{2}^{*}$.
In the next step, based on the $\mathcal{K}_{1}^{\ast}$ and $\mathcal{K}_{2}^{\ast}$
we jointly optimize the other network parameters. 

\subsection{Optimal SRUS }

The SRUS protocol must address two questions:
\begin{itemize}
\item How do we seleect the relaying users?
\item How many relaying users are needed?
\end{itemize}
In order to address these research questions, we turn to the following
proposition.
\begin{prop}
\label{prop:At-the-global}At the global by optimal point $\left(\mathbf{P}^{*},\theta^{*},\mathbf{\mathbf{\boldsymbol{\chi}}}^{*},\mathcal{K}_{1}^{*},\left\{ p_{j}^{*}\right\} _{\forall j\in\mathcal{K}_{1}^{*}},p_{z}^{*}\right)$
of problem \eqref{eq:18}, the common secrecy rates achieved by the
users in the set $\mathcal{K}_{1}^{*}$, i.e., $\mathcal{R}_{c,\mathcal{K}_{1}^{*}}^{sec}$,
and set $\mathcal{K}_{2}^{*}$, i.e., $\mathcal{R}_{c,\mathcal{K}_{2}^{*}}^{sec}$,
are equal which can be formulated as:
\begin{gather}
\underset{\mathcal{R}_{c,\mathcal{K}_{1}^{*}}^{\text{sec}}}{\underbrace{\underset{k\in\mathcal{K}_{1}^{*}}{\min}\left\{ \mathcal{R}_{c,k}^{\text{sec}\,(1)}\right\} }}=\underset{\mathcal{R}_{c,\mathcal{K}_{2}^{*}}^{\text{sec}}}{\underbrace{\underset{k\in\mathcal{K}_{2}^{*}}{\min}\left\{ \mathcal{R}_{c,k}^{\text{sec}\,(1)}+\mathcal{R}_{c,k}^{\text{sec}\,(2)}\right\} }},\label{eq:19}
\end{gather}
\end{prop}
\begin{IEEEproof}
See Appendix A.
\end{IEEEproof}
\begin{rem}
It can be an be readily concluded from Proposition \ref{prop:At-the-global}
that the optimal secure relaying user grouping obeys the following
rule, when $0<\theta^{*}<1$:
\begin{equation}
\underset{k\in\mathcal{K}_{1}^{*}}{\min}\left\{ \mathcal{R}_{c,k}^{\text{sec}\,(1)}\right\} >\underset{k\in\mathcal{K}_{2}^{*}}{\min}\left\{ \mathcal{R}_{c,k}^{\text{sec}\,(\text{2})}\right\} .\label{eq:21-1}
\end{equation}
\begin{equation}
\underset{k\in\mathcal{K}_{1}^{*}}{\min}\left\{ R_{c,k}^{(1)}\right\} >\underset{k\in\mathcal{K}_{2}^{*}}{\min}\left\{ R_{c,k}^{(\text{1})}\right\} .\label{eq:20}
\end{equation}
\end{rem}
Given the insight inferred from Proposition \ref{prop:At-the-global},
to enhance the common secrecy rate, the users having larger $\mathcal{R}_{c,k}^{\text{sec}\,(1)}$
in BP and smaller RP leakage tend to be clustered in $\mathcal{K}_{1}$,
while the users with lower $\mathcal{R}_{c,k}^{\text{sec}\,(1)}$
in BP and larger RP leakage tend to fall into $\mathcal{K}_{2}$.
Since the $\mathcal{R}_{c,k}^{\text{sec}\,(1)}$ and RP leakage respectively
depends on $\left\Vert {\mathbf{h}_{k}}\right\Vert _{2}$ and $\left|\hat{h}_{k,e}\right|$,
an intuitive and simple selection algorithm is based on the simple
metric $\frac{\left\Vert {\mathbf{h}_{k}}\right\Vert _{2}}{\left|\hat{h}_{k,e}\right|}$.
In the following a pair of relaying protocols (centralized and distributed)
based on this metric are presented.
\begin{itemize}
\item \textbf{Centralized relaying protocol}: Since the UAV-BS needs all
the CSIs for its TPC design, one option is to perform SRUS by the
UAV-BS. The proposed centralized SRUS algorithm is presented in Algorithm
1, where the process of channel estimation is performed through the
classic Request-To-Send (RTS)/Clear-To-Send (CTS) collision avoidance
mechanism.
\item \textbf{Distributed relaying protocol:} In the above centralized protocol,
the RTS packet is transmitted through a common downlink pilot channel,
while the CTS packets are fed back through individual uplink pilot
channels dedicated to each user. Because of this difference between
the uplink and downlink centralized techniques are more susceptible
to channel imperfections and thus, distributed selection is preferred.
More explicitly, while the centralized protocol needs the CSIT for
all users at the UAV-BS to select the best secure relaying users,
the distributed protocol allows users to select their secure relaying
partners based on the CSI estimated at the Receiver (CSIR). This may
be readily obtained from the common downlink pilot channels. The proposed
distributed SRUS algorithm is presented in Algorithm 2.
\end{itemize}

\section{Network Parameter Optimization\label{sec:IV}}

\begin{algorithm}[tbh]
\caption{$\textbf{Centralized SRUS Protocol}$}
$\textbf{Initialization }$ $n=0,\,$$\mathcal{K}_{1}=\textrm{Ø}$,
and $\mathcal{K}_{2}=\mathcal{K}$.

$\textbf{Step 1 }$ Channel estimation of Users at UAV-BS: First UAV-BS
inform all users about $\left|\mathcal{K}_{1}\right|=K_{1}$ through
a RTS packet and then they respond through a CTS packet by which their
channel is estimated at UAV-BS.

$\textbf{Step 2 }$ Ordering the users based on the channel strength
$\frac{\left\Vert {\mathbf{h}_{k}}\right\Vert _{2}}{\left|\hat{h}_{k,e}\right|}$.

$\textbf{Step 3 }$ UAV-BS finds $\mathcal{K}_{1}$ secure relaying
user as follows.

~Repeat:\vspace{0.1cm}

\begin{tabular}{c|c}
$\text{\ding{202}}$ & $k^{\star}=\underset{k}{\max}\left\{ \frac{\left\Vert {\mathbf{h}_{k}}\right\Vert _{2}}{\left|\hat{h}_{k,e}\right|}\right\} $\tabularnewline
$\text{\ding{183}}$ & $\mathcal{K}_{1}\longleftarrow\mathcal{K}_{1}\cup\left\{ k^{\star}\right\} $\tabularnewline
\ding{184} & $\mathcal{K}_{2}\longleftarrow\mathcal{K}_{2}\backslash\left\{ k^{\star}\right\} $\tabularnewline
\end{tabular}\vspace{0.1cm}

$\,\mathrm{Until}$ $\left|\mathcal{K}_{1}\right|=K_{1}$.

$\textbf{Output:}$ UAV-BS transmits a \textquotedbl\textit{flag}\textquotedbl{}
packet containing the selection results to all users.
\end{algorithm}
\begin{algorithm}[tbh]
\caption{$\textbf{Decentralized SRUS Protocol}$}
$\textbf{Initialization }$ $n=0,\,$$\mathcal{K}_{1}=\textrm{Ø}$,
and $\mathcal{K}_{2}=\mathcal{K}$.

$\textbf{Step 1 }$ UAV-BS broadcasts the $\left|\hat{h}_{k,e}\right|$
and RTS packet that contains the value of $\left|\mathcal{K}_{1}\right|=K_{1}$
to all users. Each user receives them, replies to UAV-BS with CTS
and estimates the channel ${\mathbf{h}_{k}}$ and calculates the $\frac{\left\Vert {\mathbf{h}_{k}}\right\Vert _{2}}{\left|\hat{h}_{k,e}\right|}$.

$\textbf{Step 2 }$ Decentralized SRUS:

~Repeat:\vspace{0.1cm}

\begin{tabular}{c|>{\raggedright}p{8cm}}
$\text{\ding{202}}$ & At each user in $\mathcal{K}_{2}$, clear the existing timer if it
has one.\tabularnewline
$\text{\ding{183}}$ & The timer is reset at each user in $\mathcal{K}_{2}$ as $t_{k}=\frac{1}{\left\Vert {\mathbf{h}_{k}}\right\Vert _{2}},\,$
$\forall k\in\mathcal{K}_{2}$.\tabularnewline
\ding{184} & The user who counts down to zero announces himself as the best secure
relaying user in $\mathcal{K}_{2}$ with a \textquotedblleft \textit{flag}\textquotedblright{}
packet.\tabularnewline
\ding{185} & Upon receiving the \textquotedblleft \textit{flag}\textquotedblright{}
packet from the best secure relaying user $k^{\star}$, each user
in $\mathcal{K}_{2}$ updates itself as $\mathcal{K}_{1}\longleftarrow\mathcal{K}_{1}\cup\left\{ k^{\star}\right\} $,
$\mathcal{K}_{2}\longleftarrow\mathcal{K}_{2}\backslash\left\{ k^{\star}\right\} $.\tabularnewline
\end{tabular}\vspace{0.1cm}

$\,\mathrm{Until}$ $\left|\mathcal{K}_{1}\right|=K_{1}$.
\end{algorithm}
After relaxing the problem from the discontinuous variables $\mathcal{K}_{1}^{\ast}$
and $\mathcal{K}_{2}^{\ast}$, in this stage, we aim for optimizing
the remaining variables comprised in the set $\left\{ \mathbf{P},\ \theta,\mathbf{\mathbf{\boldsymbol{\chi}}},\left\{ p_{j}\right\} _{j\in\mathcal{K}_{1}},p_{z}\ \right\} $.
However, the resultant problem is still non-convex because of the
non-convex Objective Function (OF) as well as non-convex constraint
set, and thus finding the global optimum is intractable. To circumvent
the non-convexity, we design a sub-optimal algorithm relying on the
classic SPCA \cite{38}. Using SPCA, the problem is iteratively approximated
by a sequence of  programs. To facilitate the design of SPCA, we first
introduce the auxiliary variable $r_{sec}$ and recast problem \eqref{eq:18}
into its equivalent epigraph form, given by:
\begin{equation}
\begin{array}{cc}
\underset{\mathbf{P},\theta,\mathbf{\mathbf{\boldsymbol{\chi}}},\mathcal{K}_{1},\left\{ p_{j}\right\} _{\forall j\in\mathcal{K}_{1}},p_{z}}{\max} & r_{sec}\end{array}\,\,\label{eq:21}
\end{equation}
s.t.$\textit{ }$

\begin{tabular}{>{\raggedright}p{15.7cm}}
$C_{1}:\,\,\chi_{k}\left[R_{c}-R_{c,e}\right]^{+}+\left[R_{p,k}-R_{k,e}\right]^{+}\geq r_{sec},\,\,$

$\forall k\in\mathcal{K},\,\,\,\forall\mathbf{\Delta n}_{e}\in\varTheta_{e},\,\forall\left\{ \triangle n_{j,e}\right\} _{\forall j\in\mathcal{K}_{1}}\in\varTheta_{h_{j,e}}$\tabularnewline
\end{tabular}

\begin{tabular}{>{\raggedright}p{19cm}}
(\eqref{eq:18} -$C_{1}$), (\eqref{eq:18} -$C_{2}$), (\eqref{eq:18}
-$C_{3}$),(\eqref{eq:18} -$C_{4}$), (\eqref{eq:18} -$C_{5}$),

(\eqref{eq:18} -$C_{6}$), (\eqref{eq:18} -$C_{7}$).\vspace{0.1cm}
\tabularnewline
\end{tabular}

It is easy to show that \eqref{eq:18} and \eqref{eq:21} are equivalent,
because, upon observing \eqref{eq:21}, it can be noted that $r_{sec}$
plays the role of lower bound for $\underset{\mathbf{\Delta n}_{e},\left\{ \triangle n_{j,e}\right\} _{\forall j\in\mathcal{K}_{1}}}{\max}{\mathcal{R}_{k}^{\textrm{sec}}}$
in the OF of \eqref{eq:18} and its maximization will increase the
left-side of the constraints (\ref{eq:21} -$C_{1}$), so that it
would be active at the optimum. After this transformation, the problem
\eqref{eq:21} is still non- due to the constraints (\eqref{eq:18}-$C_{1}$)-(\eqref{eq:18}-$C_{3}$).
In order to facilitate the procedure of convexifying \eqref{eq:21},
we further define new sets of auxiliary variables $\boldsymbol{\mathbf{\beta}}_{p}=\left[\beta_{p,k}\right]_{\forall k\in\mathcal{K}}$,
$\boldsymbol{\mathbf{\beta}}_{c}^{(1)}=\left[\beta_{c,k}^{(1)}\right]_{\forall k\in\mathcal{K}}$,
$\boldsymbol{\mathbf{\beta}}_{c}^{(2)}=\left[\beta_{c,k}^{(2)}\right]_{\forall k\in\mathcal{K}_{2}}$,$\boldsymbol{\mathbf{\beta}}_{e}=\left[\beta_{k,e}\right]_{\forall k\in\mathcal{K}},$$\alpha_{c,e}$,
$\boldsymbol{\mathbf{\mathbf{\rho}}}_{c}^{(1)}=\left[\rho_{c,k}^{(1)}\right]_{\forall k\in\mathcal{K}}$,
$\boldsymbol{\mathbf{\mathbf{\rho}}}_{c}^{(2)}=\left[\rho_{c,k}^{(2)}\right]_{\forall k\in\mathcal{K}_{2}}$,
and $\boldsymbol{\mathbf{\mathbf{\rho}}}_{p}=\left[\rho_{p,k}\right]_{\forall k\in\mathcal{K}}$.
$\boldsymbol{\mathbf{\beta}}_{p}$, $\boldsymbol{\mathbf{\beta}}_{c}$,
are adopted respectively for representing the rate vectors of the
private streams and the common streams (without $\theta$) at all
users , while $\alpha_{c,e}$ and $\boldsymbol{\mathbf{\beta}}_{e}$
represent the rate of the common streams and private streams at $Eve$,
respectively. Finally, $\boldsymbol{\mathbf{\mathbf{\rho}}}_{c}$
and $\boldsymbol{\mathbf{\mathbf{\rho}}}_{p}$ respectively denote
the SINR vectors of the private streams and the common streams at
all the users. This allows us to reformulate \eqref{eq:21} into:
\begin{equation}
\begin{array}{cc}
\underset{\mathbf{P},\theta,\mathbf{\mathbf{\boldsymbol{\chi}}},\mathcal{K}_{1},\left\{ p_{j}\right\} _{\forall j\in\mathcal{K}_{1}},p_{z},\boldsymbol{\mathbf{\beta}}_{\left\{ c,e,p\right\} },\boldsymbol{\mathbf{\mathbf{\rho}}}_{\left\{ c,p\right\} },\alpha_{c,e}}{\max} & r_{sec}\end{array}\,\,\label{eq:22}
\end{equation}
s.t.$\textit{ }$

\begin{tabular}{>{\raggedright}p{8cm}}
$C_{1}:\,\,x_{k}\left(R_{c}-\alpha_{c,e}\right)+\theta\left(\alpha_{p,k}-\alpha_{e,k}\right)\geq r_{sec},\,\,\forall k\in\mathcal{K}$\tabularnewline
\end{tabular}

\begin{tabular}{>{\raggedright}p{15.7cm}}
$C_{2}:\,\,\theta\alpha_{c,j}\geq R_{c},\,\,\forall j\in\mathcal{K}_{1}$\tabularnewline
\end{tabular}

\begin{tabular}{>{\raggedright}p{15.7cm}}
$C_{3}:\,\,\theta\alpha_{c,k}+R_{c,k}^{(2)}\geq R_{c},\,\,\forall k\in\mathcal{K}_{2}$\tabularnewline
\end{tabular}

\begin{tabular}{>{\raggedright}p{15.7cm}}
$C_{4}:\,\,\frac{\left|{\mathbf{h}_{k}}^{H}\mathbf{p}_{k}\right|^{2}}{\underset{j\in\mathcal{K},j\neq k}{\sum}\left|{\mathbf{h}_{k}}^{H}\mathbf{p}_{j}\right|^{2}+\sigma_{k}^{2}}\geq\rho_{p,k},\,\,\forall k\in\mathcal{K}$\tabularnewline
\end{tabular}

\begin{tabular}{>{\raggedright}p{15.7cm}}
$C_{5}:\,\,\frac{\left|{\mathbf{h}_{k}}^{H}\mathbf{p}_{c}\right|^{2}}{\underset{j\in\mathcal{K}}{\sum}\left|{\mathbf{h}_{k}}^{H}\mathbf{p}_{j}\right|^{2}+\sigma_{k}^{2}}\geq\rho_{c,k}^{(1)},\,\,\forall k\in\mathcal{K}$\tabularnewline
\end{tabular}

\begin{tabular}{>{\raggedright}p{15.7cm}}
$C_{6}:\,\,1+\rho_{c,k}^{(1)}-2^{\beta_{c,k}^{(1)}}\geq0,\,\,\forall k\in\mathcal{K}$\tabularnewline
\end{tabular}

\begin{tabular}{>{\raggedright}p{15.7cm}}
$C_{7}:\,\,1+\rho_{p,k}-2^{\beta_{p,k}}\geq0,\,\,\forall k\in\mathcal{K}$\tabularnewline
\end{tabular}

\begin{tabular}{>{\raggedright}p{15.7cm}}
$C_{8}:\,\,R_{c}\geq r_{c},$\tabularnewline
\end{tabular}

\begin{tabular}{>{\raggedright}p{15.7cm}}
$C_{9}:\,\,R_{c}\geq\alpha_{c,e},\,\,\beta_{p,k}\geq\beta_{k,e},\,\,\forall k\in\mathcal{K}$\tabularnewline
\end{tabular}

\begin{tabular}{>{\raggedright}p{15.7cm}}
$C_{10}:$ $r_{c}\geq\underset{\mathbf{\Delta n}_{e},\left\{ \triangle n_{j,e}\right\} _{\forall j\in\mathcal{K}_{1}}}{\max}{R_{c,e}}$,\tabularnewline
\end{tabular}

\begin{tabular}{>{\raggedright}p{15.7cm}}
$C_{11}:\,\,\alpha_{c,e}\geq\underset{\mathbf{\Delta n}_{e},\left\{ \triangle n_{j,e}\right\} _{\forall j\in\mathcal{K}_{1}}}{\max}{R_{c,e}}$\tabularnewline
\end{tabular}

\begin{tabular}{>{\raggedright}p{15.7cm}}
$C_{12}:\,\,\alpha_{k,e}\geq\underset{\mathbf{\Delta n}_{e}}{\max}{R_{k,e}}$$,\,\,\alpha_{k,e}\geq\theta\beta_{k,e}$\tabularnewline
\end{tabular}

\begin{tabular}{>{\raggedright}p{15.7cm}}
$C_{13}:\,\,\underset{k\in\mathcal{K}}{\sum}x_{k}=1$,$\,0\leq x_{k}\leq1,\,\,\forall k\in\mathcal{K}$.\tabularnewline
\end{tabular}

Despite this linearization, by invoking the definitions of rates,
the constraints (\ref{eq:22}-$C_{1}:C_{5}$) and (\ref{eq:22}-$C_{10}:C_{12}$)
are still non-convex. To handle the non-convexity of these constraints
we construct a suitable convex inner subset for approximating the
non-convex feasible solution set. Given this perspective, we first
try to circumvent the bilinear factor $\theta\beta_{p,k}$ that appeared
in (\ref{eq:22}-$C_{1}:C_{3}$) which can be equivalently reformulated
with the aim of linearization as $\theta\beta_{p,k}$ = $\frac{1}{4}\left(\theta+\beta_{p,k}\right)^{2}-\frac{1}{4}\left(\theta-\beta_{p,k}\right)^{2}$.
Using its first-order Taylor expansion counterpart, at the $m$-th
iteration, $\theta\beta_{p,k}$ is approximated at the point $\left(\theta^{[m]},\beta_{p,k}^{[m]}\right)$
as follows:
\begin{equation}
\theta\beta_{p,k}\geq\varTheta^{[m]}\left(\theta,\beta_{p,k}\right),\label{eq:23}
\end{equation}
\begin{gather*}
\varTheta^{[m]}\left(\theta,\beta_{p,k}\right)\triangleq\frac{1}{2}\left(\theta^{[m]}+\beta_{p,k}^{[m]}\right)\left(\theta+\beta_{p,k}\right)-\frac{1}{4}\left(\theta^{[m]}+\beta_{p,k}^{[m]}\right)^{2}\\
-\frac{1}{4}\left(\theta-\beta_{p,k}\right)^{2}.
\end{gather*}

Similarly, to acquire the lower bound of $\theta\beta_{k,e}$, we
approximate the term $(\theta-\beta_{k,e})^{2}$, which appeared in
its expanded form, around $\left(\theta^{[m]},\beta_{k,e}^{[m]}\right)$
as follows:
\begin{equation}
\theta\beta_{k,e}\leq\bar{\varTheta}^{[m]}\left(\theta,\beta_{k,e}\right),\label{eq:24}
\end{equation}
\begin{gather*}
\bar{\varTheta}^{[m]}\left(\theta,\beta_{k,e}\right)\triangleq\frac{1}{4}(\theta+\beta_{k,e})^{2}+\frac{1}{4}(\theta^{[m]}-\beta_{k,e}^{[m]})^{2}\\
-\frac{1}{2}(\theta^{[m]}-\beta_{k,e}^{[m]})\left(\theta-\beta_{k,e}\right).
\end{gather*}

By substituting the affine approximation terms obtained in \eqref{eq:23}
and \eqref{eq:24}, the constraints (\ref{eq:22}-$C_{1}:C_{3}$)
around the point $\left(\theta^{[m]},\beta_{p,k}^{[m]},\beta_{k,e}^{[m]},\beta_{c,j}^{(1)^{[m]}}\right)$
are approximated as follows:

\begin{equation}
C_{k}-\alpha_{c,e}+\varTheta^{[m]}\left(\theta,\beta_{p,k}\right)-\bar{\varTheta}^{[m]}\left(\theta,\beta_{k,e}\right)\geq r_{sec},\,\,\,\forall k\in\mathcal{K}\text{,}\label{eq:25}
\end{equation}
\begin{equation}
\varTheta^{[m]}\left(\theta,\beta_{c,k}^{(1)}\right)\geq\underset{\forall k^{'}\in\mathcal{K}}{\sum}C_{k^{'}},\,\,\forall k\in\mathcal{K}_{1}\text{,}\label{eq:26}
\end{equation}
\begin{equation}
\varTheta^{[m]}\left(\theta,\beta_{c,k}^{(1)}\right)+R_{c,k}^{(2)}\geq\underset{\forall k^{'}\in\mathcal{K}}{\sum}C_{k^{'}},\,\,\forall k\in\mathcal{K}_{2}.\label{eq:27}
\end{equation}

As for the constraints (\ref{eq:22}-$C_{4}:C_{5}$), they can be
equivalently expressed through the Difference-of-Convex (DC) decomposition
\cite{34}, given by:
\begin{equation}
\underset{j\in\mathcal{K},j\neq k}{\sum}\left|{\mathbf{h}_{k}}^{H}\mathbf{p}_{j}\right|^{2}+\sigma_{k}^{2}-\underset{\mathcal{B}_{p,k}}{\underbrace{\frac{\left|{\mathbf{h}_{k}}^{H}\mathbf{p}_{k}\right|^{2}}{\rho_{p,k}}}}\leq0,\,\,\forall k\in\mathcal{K}\label{eq:28}
\end{equation}
\begin{equation}
\underset{j\in\mathcal{K}}{\sum}\left|{\mathbf{h}_{k}}^{H}\mathbf{p}_{j}\right|^{2}+\sigma_{k}^{2}-\underset{\mathcal{C}_{p,k}}{\underbrace{\frac{\left|{\mathbf{h}_{k}}^{H}\mathbf{p}_{c}\right|^{2}}{\rho_{c,k}^{(1)}}}}\leq0,\,\,\forall k\in\mathcal{K}.\label{eq:29}
\end{equation}

As it can be observed, the non-convexities of \eqref{eq:28} and \eqref{eq:29}
are caused by the concave terms $\mathcal{B}_{p,k}$ and $\mathcal{C}_{p,k}$.
Therefore, we replace them by their affine approximation counterparts
obtained by the first-order Taylor expansion around the point $\left({\mathbf{p}_{c}}^{[m]},{\mathbf{p}_{k}}^{[m]},{\rho_{c,k}^{(1)}}^{[m]},{\rho_{p,k}}^{[m]}\right)$
and obtain the convex approximations of \eqref{eq:28} and \eqref{eq:29}
as follows:
\begin{equation}
\underset{j\in\mathcal{K},j\neq k}{\sum}\left|{\mathbf{h}_{k}}^{H}\mathbf{p}_{j}\right|^{2}+\sigma_{k}^{2}-\Psi^{[m]}\left({\mathbf{p}_{k}},{\rho_{p,k}};\,{\mathbf{h}_{k}}\right)\leq0,\,\,\forall k\in\mathcal{K}\label{eq:30}
\end{equation}
\begin{equation}
\underset{j\in\mathcal{K}}{\sum}\left|{\mathbf{h}_{k}}^{H}\mathbf{p}_{j}\right|^{2}+\sigma_{k}^{2}-\Psi^{[m]}\left({\mathbf{p}_{c}},{\rho_{c,k}}^{(1)};\,{\mathbf{h}_{k}}\right)\leq0,\,\,\forall k\in\mathcal{K}\label{eq:31}
\end{equation}
where 
\begin{equation}
\Psi^{[m]}\left(\mathbf{u},{x};\,{\mathbf{h}}\right)\triangleq\frac{2\Re e\left\{ \left({{\mathbf{u}}^{[m]}}\right)^{H}{\mathbf{h}}\mathbf{h}^{H}\mathbf{u}\right\} }{{x}^{[m]}}-\frac{\left|\mathbf{h}^{H}{\mathbf{u}}^{[m]}\right|^{2}x}{\left({x}^{[m]}\right)^{2}}.\label{eq:32}
\end{equation}

Now, with the objective of convexifying (\ref{eq:22}-$C_{10}:C_{11}$),
we can equivalently write them as follows:
\begin{equation}
r_{c}\geq\alpha_{c,e},\label{eq:33}
\end{equation}
\begin{equation}
\alpha_{c,e}\geq R_{c,e},\,\,\,\forall\mathbf{\Delta n}_{e}\in\varTheta_{e},\,\forall\left\{ \triangle n_{j,e}\in\varTheta_{h_{j,e}}\right\} _{\forall j\in\mathcal{K}_{1}}.\label{eq:34}
\end{equation}

However, \eqref{eq:34} is still non-convex, which enforces us to
introduce the new auxiliary variables $\left\{ {\rho_{c,e}^{(1)},\rho_{c,e}^{(2)},\beta_{c,e}^{(1)},\beta_{c,e}^{(2)}}\right\} $
representing the SINR, as well as the rate of the common streams associated
with first and second phases at $Eve$, respectively. After some routine
mathematical manipulations, \eqref{eq:34} can be recast as: 
\begin{equation}
\bar{\varTheta}^{[m]}\left(\theta,\beta_{c,e}^{(1)}\right)+\bar{\varTheta}^{[m]}\left(1-\theta,\beta_{c,e}^{(2)}\right)\text{\ensuremath{\le}}\,\alpha_{c,e},\label{eq:35}
\end{equation}
\begin{equation}
1+\rho_{c,e}^{(j)}-\varGamma^{[m]}\left(\beta_{c,e}^{(j)}\right)\leq0,\,\,j\in\left\{ 1,2\right\} ,\label{eq:36}
\end{equation}
\begin{equation}
\frac{\left|{\mathbf{h}_{e}}^{H}\mathbf{p}_{c}\right|^{2}}{\underset{k\in\mathcal{K}}{\sum}\left|{\mathbf{h}_{e}}^{H}\mathbf{p}_{k}\right|^{2}+\sigma_{e}^{2}}\leq\rho_{c,e}^{(1)},\,\,\,\forall\mathbf{\Delta n}_{e}\in\varTheta_{e},\label{eq:37}
\end{equation}
\begin{equation}
\frac{\underset{j\in\mathcal{K}_{1}}{\sum}p_{j}\left|{h_{j,e}}\right|^{2}}{p_{z}\left\Vert {\mathbf{h}_{e}}^{H}\mathbf{\hat{p}}_{z}\right\Vert ^{2}+\sigma_{e}^{2}}\leq\rho_{c,e}^{(2)},\,\,\,\forall\mathbf{\Delta n}_{e}\in\varTheta_{e},\,\forall\left\{ \triangle n_{j,e}\in\varTheta_{h_{j,e}}\right\} ,\label{eq:38}
\end{equation}
where $\varGamma^{[m]}\left(x\right)\triangleq2^{x^{[m]}}\left[1+\ln(2)\left(x-x^{[m]}\right)\right].$
Here, \eqref{eq:37} and \eqref{eq:38} are still non-convex, which
enforces us to define the new auxiliary variables $\left\{ {x_{c,e},\left\{ u_{k,c,e}\right\} _{k\in\mathcal{K}},\left\{ y_{j,c,e}\right\} _{j\in\mathcal{K}_{1}},v_{c,e}}\right\} $,
by which we can reformulate \eqref{eq:37} and \eqref{eq:38} as follows:
\begin{equation}
\frac{x_{c,e}^{2}}{\underset{k\in\mathcal{K}}{\sum}u_{k,c,e}+\sigma_{e}^{2}}\leq\rho_{c,e}^{(1)},\label{eq:39}
\end{equation}
\begin{equation}
\frac{\underset{j\in\mathcal{K}_{1}}{\sum}y_{j,c,e}^{2}}{v_{c,e}+\sigma_{e}^{2}}\leq\rho_{c,e}^{(2)},\label{eq:40}
\end{equation}
\begin{equation}
\underset{\forall\mathbf{\Delta n}_{e}\in\varTheta_{e}}{\max}\left|\left({\hat{\mathbf{h}}_{e}}+{\mathbf{\Delta n}_{e}}\right)^{H}\mathbf{p}_{c}\right|\leq x_{c,e},\,\label{eq:41}
\end{equation}
\begin{equation}
\underset{\forall\mathbf{\Delta n}_{e}\in\varTheta_{e}}{\min}\left|\left({\hat{\mathbf{h}}_{e}}+{\mathbf{\Delta n}_{e}}\right)^{H}\mathbf{p}_{k}\right|^{2}\geq u_{k,c,e},\,\,\forall k\in\mathcal{K},\label{eq:42}
\end{equation}
\begin{equation}
\underset{\triangle n_{j,e}\in\varTheta_{h_{j,e}}}{\max}\delta_{j}\left|\hat{h}_{j,e}+\triangle n_{j,e}\right|\leq y_{j,c,e},\,\,\forall j\in\mathcal{K}_{1},\label{eq:43}
\end{equation}
\begin{equation}
\underset{\forall\mathbf{\Delta n}_{e}\in\varTheta_{e}}{\min}p_{z}\left|\left({\hat{\mathbf{h}}_{e}}+{\mathbf{\Delta n}_{e}}\right)^{H}\mathbf{\hat{p}}_{z}\right|^{2}\geq v_{c,e},\label{eq:44}
\end{equation}
where $\delta_{j}\triangleq\sqrt{p_{j}}$. Next, we adopt the SPCA
method to convert the non-convex constraints \eqref{eq:39}-\eqref{eq:40}
into convex constraints. In this regard, we introduce the auxiliary
variables $\left\{ d_{c,e}^{(1)},d_{c,e}^{(2)}\right\} $ into the
constraints \eqref{eq:39}-\eqref{eq:40}, leading to:
\begin{equation}
\underset{k\in\mathcal{K}}{\sum}u_{k,c,e}+\sigma_{e}^{2}\geq d_{c,e}^{(1)},\label{eq:45}
\end{equation}
\begin{equation}
\frac{x_{c,e}^{2}}{d_{c,e}^{(1)}}\leq\rho_{c,e}^{(1)},\label{eq:46}
\end{equation}
\begin{equation}
v_{c,e}+\sigma_{e}^{2}\geq d_{c,e}^{(2)},\label{eq:47}
\end{equation}
\begin{equation}
\frac{\underset{j\in\mathcal{K}_{1}}{\sum}y_{j,c,e}^{2}}{d_{c,e}^{(2)}}\leq\rho_{c,e}^{(2)}.\label{eq:48}
\end{equation}

Based on the above approximation methods, the original optimization
problem can be solved using the SPCA method. The main idea of SPCA
is to successively solve a sequence of convex sub-problems. At the
$m$-th iteration, based on the optimal solution $\left(\mathbf{P}^{[m]},\theta^{[m]},\mathbf{\mathbf{\boldsymbol{\chi}}}^{[m]},\left\{ p_{j}^{[m]}\right\} _{\forall j\in\mathcal{K}_{1}},p_{z}^{[m]},\boldsymbol{\mathbf{\beta}}_{\left\{ c,e,p\right\} }^{[m]},\boldsymbol{\mathbf{\mathbf{\rho}}}_{\left\{ c,p\right\} }^{[m]},\alpha_{c,e}^{[m]}\right)$
obtained from the previous $\left(m-1\right)$-st iteration, we solve
the following sub-problem:
\begin{multline}
\begin{array}{cc}
\underset{\begin{array}{c}
\mathbf{P},\theta,\mathbf{\mathbf{\boldsymbol{\chi}}},\mathcal{K}_{1},\left\{ p_{j}\right\} _{\forall j\in\mathcal{K}_{1}},\\
p_{z},\boldsymbol{\mathbf{\beta}}_{\left\{ c,e,p\right\} },\boldsymbol{\mathbf{\mathbf{\rho}}}_{\left\{ c,p\right\} },\alpha_{c,e}
\end{array}}{\max} & r_{sec}^{[m]}\end{array}\,\,\label{eq:22-1}
\end{multline}
s.t.$\textit{ }$

\begin{tabular}{>{\raggedright}p{9cm}}
\eqref{eq:23}-\eqref{eq:27}, \eqref{eq:31}, \eqref{eq:32}, \eqref{eq:36},
\eqref{eq:37}, \eqref{eq:46}-\eqref{eq:49}, \ref{eq:22}-$C_{13}$,
(\ref{eq:22} -$C_{4}:C_{7}$), (\ref{eq:18} -$C_{4}:C_{7}$).\tabularnewline
\end{tabular}

\noindent Using the interior-point methods of \cite{34}, we can solve
the problem in \eqref{eq:21-1}, which is a convex Quadratically Constrained
Quadratic Program (QCQP) \cite{34}.

\subsection{Feasible Initial Point Search Algorithm }

Note that, if \eqref{eq:22-1} is initialized by random points, it
may fail at the very beginning, because of infeasibility \cite{1,5,10}.
To circumvent this issue, we now conceive a feasible initial point
search algorithm (FIPSA) in this section. In this regard, we aim for
minimizing an infeasibility indicator parameter $\vartheta>0$, to
flag up any violation of the constraints of \eqref{eq:22-1}. Hence
we have to rewrite all the constraints of problem \eqref{eq:22-1}
in the form of $\left.\mathcal{\text{\ensuremath{\mathscr{G}}}}_{i}\left(\mathbf{x}\right)\right|_{i=1}^{16}\le\vartheta$,
where $\begin{array}{c}
\mathbf{x}\triangleq\left(\mathbf{P},\theta,\mathbf{\mathbf{\boldsymbol{\chi}}},\mathcal{K}_{1},\left\{ p_{j}\right\} _{\forall j\in\mathcal{K}_{1}},p_{z},\boldsymbol{\mathbf{\beta}}_{\left\{ c,e,p\right\} },\boldsymbol{\mathbf{\mathbf{\rho}}}_{\left\{ c,p\right\} },\alpha_{c,e}\right)\end{array}$ and $\mathcal{\text{\ensuremath{\mathscr{G}}}}_{i}\left(\mathbf{x}\right)$
stands for the reshaped format of the $i$-th constraint and all the
terms at the left-side of the less than or equal to zero. Then we
reformulate the feasibility problem as follows: 
\begin{equation}
\begin{array}{cc}
\underset{\mathbf{x}}{\min} & \vartheta\end{array}\,\,\label{eq:38-1}
\end{equation}
s.t.$\textit{ }$%
\begin{tabular}{>{\raggedright}p{15.7cm}}
$\left.\mathcal{\text{\ensuremath{\mathscr{G}}}}_{i}\left(\mathbf{x}\right)\right|_{i=1}^{16}\,\text{\ensuremath{\le\,}}\vartheta.$\vspace{0.05cm}
\tabularnewline
\end{tabular}

This approach has been previously proposed in \cite{9,12} as a low-complexity
technique of finding feasible initial points. Overall, the proposed
FIPSA runs at the first step and then the initial points (IP)s calculated
are fed to \eqref{eq:22-1}. As a starting point, FIPSA commences
with following IPs and the algorithm is halted if either the stopping
criterion is satisfied or the maximum number of affordable iterations
is reached. In this regard, the TPCs of the proposed FIPSA algorithm
are initialized by using MRT combined with Singular Value Decomposition
(SVD). The TPC $\left\{ \mathbf{p}_{k}^{[0]}\right\} _{\forall k\in\mathcal{K}}$
constructed for the private stream $\left\{ s_{k}\right\} _{\forall k\in\mathcal{K}}$
is initialized as $\sqrt{\omega\frac{P_{t}}{K}}\frac{{\mathbf{h}_{k}}}{\left\Vert {\mathbf{h}_{k}}\right\Vert }$,
where $0<\omega<1$. The TPC ${\mathbf{p}_{c}}^{[0]}$ for the common
message $s_{c}$ is initialized as $\sqrt{\left(1-\omega\right)P_{t}}\mathbf{u}_{c}$,
and $\mathbf{u}_{c}$ is the largest left singular vector of the channel
matrix $\mathbf{A}\triangleq\left[{\mathbf{h}_{1}},{\mathbf{h}_{2}},...,{\mathbf{h}_{K}}\right]$.
It is calculated by $\mathbf{u}_{c}\triangleq\mathbf{U}_{c}\left(:,1\right)$
where $\mathbf{A}\triangleq\mathbf{U}\mathbf{S}\mathbf{D}^{H}$. Also
we have initialized the iterative variables as $\theta^{[0]}=0.5$,
$p_{z}^{[0]}=\omega_{z}\bar{P}_{z}$, $p_{j}^{[0]}=\omega_{j}\bar{P}_{J}$
where $0<\left\{ \omega_{j},\omega_{z}\right\} <1$, $\rho_{c,k}^{(1)^{[0]}}=\frac{\left|{\mathbf{h}_{k}}^{H}\mathbf{p}_{c}^{[0]}\right|^{2}}{\underset{j\in\mathcal{K}}{\sum}\left|{\mathbf{h}_{k}}^{H}\mathbf{p}_{j}^{[0]}\right|^{2}+\sigma_{k}^{2}}$,
$\rho_{c,k}^{(2)^{[0]}}=\frac{\underset{j\in\mathcal{K}_{1}}{\sum}p_{j}^{[0]}\left|{h_{j,k}}\right|^{2}}{p_{z}^{[0]}\left\Vert {\mathbf{h}_{k}}^{H}\mathbf{\hat{p}}_{z}\right\Vert ^{2}+\sigma_{k}^{2}}$,
$\rho_{c,e}^{(1)^{[0]}}=\frac{\left|{\mathbf{h}_{e}}^{H}\mathbf{p}_{c}^{[0]}\right|^{2}}{\underset{j\in\mathcal{K}}{\sum}\left|{\mathbf{h}_{e}}^{H}\mathbf{p}_{j}^{[0]}\right|^{2}+\sigma_{e}^{2}}$,
$\rho_{c,e}^{(2)^{[0]}}=\frac{\underset{j\in\mathcal{K}_{1}}{\sum}p_{j}^{[0]}\left|{h_{j,e}}\right|^{2}}{p_{z}^{[0]}\left\Vert {\mathbf{h}_{e}}^{H}\mathbf{\hat{p}}_{z}\right\Vert ^{2}+\sigma_{e}^{2}}$,
$\rho_{p,k}^{[0]}=\frac{\left|{\mathbf{h}_{k}}^{H}\mathbf{p}_{k}^{[0]}\right|^{2}}{\underset{j\in\mathcal{K},j\neq k}{\sum}\left|{\mathbf{h}_{k}}^{H}\mathbf{p}_{j}^{[0]}\right|^{2}+\sigma_{k}^{2}}$,
$\rho_{k,e}^{[0]}=\frac{\left|{\mathbf{h}_{e}}^{H}\mathbf{p}_{k}^{[0]}\right|^{2}}{\underset{j\in\mathcal{K},j\neq k}{\sum}\left|{\mathbf{h}_{e}}^{H}\mathbf{p}_{j}^{[0}\right|^{2}+\sigma_{k}^{2}}$,
$\beta_{c,k}^{(n)^{[0]}}=\log_{2}\left(1+\rho_{c,k}^{(n)^{[0]}}\right)$,
$\beta_{c,e}^{(n)^{[0]}}=\log_{2}\left(1+\rho_{c,e}^{(n)^{[0]}}\right)\,$
$\forall n\in\left\{ 1,2\right\} ,$ $\beta_{p,k}^{[0]}=\log_{2}\left(1+\rho_{p,k}^{[0]}\right)$,
$\beta_{k,e}^{[0]}=\log_{2}\left(1+\rho_{k,e}^{[0]}\right)$, $\alpha_{c,e}^{[0]}=\theta^{[0]}\beta_{c,e}^{(1)^{[0]}}+\left(1-\theta^{[0]}\right)\beta_{c,e}^{(2)^{[0]}},$
$R_{c}^{[0]}=\min\left\{ {\left\{ \theta^{[0]}\beta_{c,j}^{(1)^{[0]}}\right\} _{j\in\mathcal{K}_{1}}},\right.\left.\left\{ \theta^{[0]}\beta_{c,k}^{(1)^{[0]}}+\left(1-\theta^{[0]}\right)\beta_{c,k}^{(2)^{[0]}}\right\} _{k\in\mathcal{K}_{2}}\right\} $
and $\chi_{k}^{[0]}=\frac{1}{K}.$

\subsection{Convergence Analysis}

The proposed SPCA-based algorithm iteratively solves the approximated
problem (\ref{eq:22}) until convergence is reached, where $\delta$
represents the convergence tolerance. In this regard, we formulate
the following proposition. 
\begin{prop}
\label{prop:For-any-feasible}The proposed SPCA-based algorithm guarantees
convergence to a stationary point of problem (\ref{eq:21}) for any
feasible initial point.
\end{prop}
\begin{IEEEproof}
See Appendix B.
\end{IEEEproof}

\subsection{Complexity Analysis}

The SPCA-based solution suggested solves the convex sub-problem \eqref{eq:22-1}
in each iteration. Problem \eqref{eq:22-1} is a generalized nonlinear
convex program due to the exponential cone constraints (\ref{eq:22}-$C_{6}:C_{7}$).
In order to approximate (\ref{eq:22}-$C_{6}:C_{7}$), a sequence
of Second Order Cones (SOCs) would be more efficient \cite{35}. Hence,
we can use one of the MATLAB optimization Toolbox solvers, such as
\textit{cvx} or \textit{fmincon}. Considering the computational complexity
of interior-point methods, i.e., $\mathcal{O}\left(\left[KN_{t}\right]^{3.5}\right)$,
it is possible to solve the resultant SOC program. The total number
of iterations required for achieving convergence may be shown to be
on the order of $\mathcal{O}\left(-\log\left(\zeta\right)\right)$.
Therefore, the worst-case computational complexity is $\mathcal{O}\left(-\log\left(\delta\right)\times\left[KN_{t}\right]^{3.5}\right)$.

\section{NUMERICAL RESULTS\label{sec:V}}

In this section, we present numerical results for characterizing our
proposed framework using the following simulation setting, unless
stated otherwise. The simulation results are averaged over $10^{2}$
random realizations of the proposed scheme. Through this section,
a maximum of $25$ iterations are considered for our proposed secrecy
fairness maximization problem to converge. Moreover, we set the maximum
convergence tolerance to be $\delta=10^{-2}$. In contrast to the
fading model between users, the path loss model of our UAV network
includes both LoS and NLoS in conjunction with the path-loss exponents
$\mathcal{L}_{n,k}^{i}=2\,$ and $\mathcal{N}_{n,k}^{i}=3.5$, respectively.
We assume that the transmit power obeys $\bar{P}_{J}=\bar{P}_{z}=\frac{P_{t}}{2},\,\,\forall k\in\mathcal{K}_{1}$
and $P_{t}=40$dBm. The UAV is assumed to serve users within a radius
of $R=300\,\textrm{m}$ at an altitude of $H=130\,\text{m}$ with
$N_{t}=4$. Furthermore, the $K=4$ users are randomly located within
the coverage area of each UAV-BS. A channel estimation error variance
of $\zeta=10^{-3}$ is assumed and the additive noise at the receivers
is considered to have a normalized power of $\sigma^{2}=\sigma_{e}^{2}=0\,\textrm{dBm}$,
and the minimum required transmission rate $r_{c}=1$. For simplicity,
we collect all the simulation parameters in table III. We compare
the following SRUS protocols for the design of $\mathcal{K}_{1}$
and $\mathcal{K}_{2}$:
\begin{table}[tbh]
\caption{List of simulation parameters}

\centering{}%
\begin{tabular}{|c|c|c|c|}
\hline 
Parametere & Value & Parametere & Value\tabularnewline
\hline 
\hline 
$\delta$ & $10^{-2}$ & $K$ & $4$\tabularnewline
\hline 
$\mathcal{L}_{n,k}^{i}$ & $2\,$ & $\zeta$ & $10^{-3}$\tabularnewline
\hline 
$\mathcal{N}_{n,k}^{i}$ & $3.5$ & $\sigma^{2}=\sigma_{e}^{2}$ & $0\,\textrm{dBm}$\tabularnewline
\hline 
$P_{t}$ & $40\,dBm$ & $r_{c}$ & $1$\tabularnewline
\hline 
$R$ & $300\,\textrm{m}$ & \#Iterations & $25$\tabularnewline
\hline 
$H$ & $130\,\text{m}$ & $N_{t}$ & $4$\tabularnewline
\hline 
\end{tabular}
\end{table}
\begin{figure}[tbh]
\begin{centering}
\includegraphics[scale=0.64]{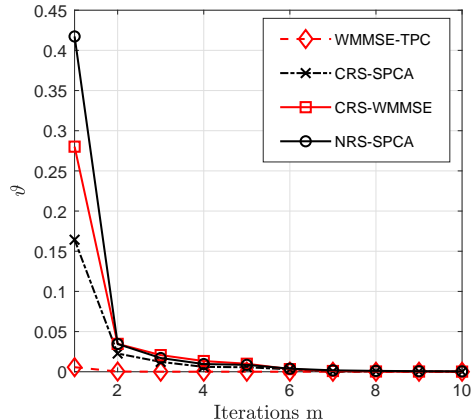}
\par\end{centering}
\caption{Evaluating of convergence speed of the proposed \textbf{FIPSA} versus
the number of iterations, $P_{t}=40$ dBm, $N_{t}=4$ and $K=4$.}
\end{figure}

\begin{itemize}
\item \textbf{Scheme 1: 1 Optimal} \textbf{Relay (1OR):} the optimal relaying
protocol where the SRUS is carried out centrally at the UAV-BS by
enumerating all possible relaying user combinations. The scheduling
scheme having the highest max-min secrecy rate is selected. It achieves
the upper bound of the max-min secrecy rate of all relaying protocols
but has the highest selection complexity.
\item \textbf{Scheme 2: 1 Best Relay (1BR):} the proposed SRUS protocols
when $\left|\mathcal{K}_{1}\right|=1$.
\item \textbf{Scheme 3: $\frac{K}{2}$ Best Relays ($\frac{k}{2}$BR}):
the proposed SRUS when $\left|\mathcal{K}_{1}\right|=\left|\mathcal{K}_{2}\right|$
. 
\item \textbf{Scheme 4: 1 Random Relay} \textbf{(1RR):} the UAV-BS randomly
selects one user from $\mathcal{K}$ and broadcasts the decision to
all users via the \textquotedblleft RTS\textquotedblright{} packet.
It has the lowest selection complexity. 
\end{itemize}
After SRUS both $\mathcal{K}_{1}$ and $\mathcal{K}_{2}$ are determined.
Then we compare the following TPC, message split and time resource
allocation algorithms: 
\begin{itemize}
\item \textbf{Algorithm 1: CRS-SPCA}: the CRS model proposed in Section
II and the proposed SPCA-based algorithm is adopted to solve problem
\eqref{eq:22-1}.
\begin{figure}[tbh]
\begin{centering}
\includegraphics[scale=0.6]{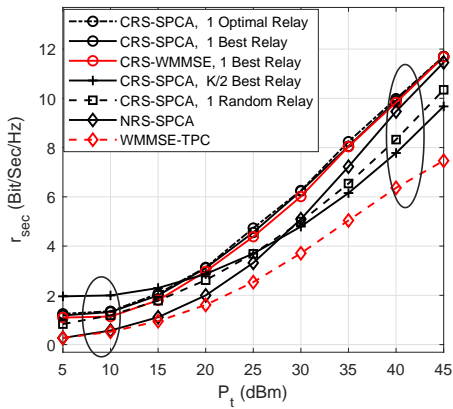}
\par\end{centering}
\caption{The $r_{sec}$ versus $P_{t}$ comparison of different strategies,
with \textbf{$\boldsymbol{N_{t}=4}$} and\textbf{ $K=4$}.}
\end{figure}
\begin{figure}[tbh]
\begin{centering}
\includegraphics[scale=0.58]{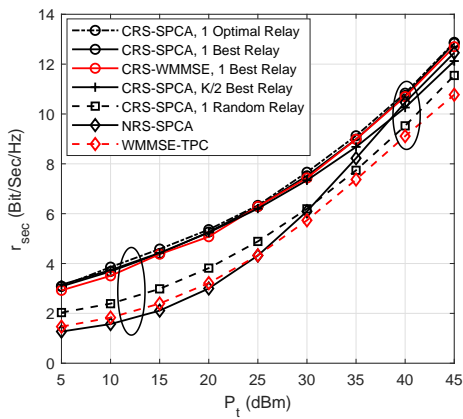}
\par\end{centering}
\caption{The $r_{sec}$ versus $P_{t}$ comparison of different strategies,
with $\boldsymbol{N_{t}=8}$ and $K=4$.}
\end{figure}
\item \textbf{Algorithm 2: CRS-WMMSE:} the CRS model proposed in Section
II, but the optimization problem \eqref{eq:22-1} is solved using
the WMMSE algorithm of \cite{17} employing one-dimensional exhaustive
search for $\theta$.
\item \textbf{Algorithm 3: NRS-SPCA}: Non-CRS is also a specific instance
of the CRS-SPCA scheme, when $\theta$ is fixed to $1$. This is the
RS scheme that has been investigated in \cite{17,31} for MISO BC
without cooperative transmission. The transmission is completed at
the end of the direct transmission phase and the cooperative transmission
is blocked.
\item \textbf{Algorithm 4: WMMSE-TPC:} the traditional multi-user linear
TPC-based beamformer investigated in \cite{33}. There is no RS and
no cooperative transmission (i.e., $\left\Vert \mathbf{p}_{c}\right\Vert =0$
and $\theta=1$).
\end{itemize}
The TPC initialization of the WMMSE algorithm is the same as that
of the proposed SPCA-based algorithm, where $\theta$ is searched
with increment $\triangle\theta=0.1$ in the CRS, WMMSE algorithm.
Therefore, the precoders and message split are optimized by using
the WMMSE algorithm $10$ times for each value of $\theta$ selected
from the set $[0.1,0.2,...,1]$. The MATLAB Toolbox \textit{cvx} is
used to solve the problem \eqref{eq:22-1}.

\begin{figure}[tbh]
\begin{centering}
\includegraphics[scale=0.56]{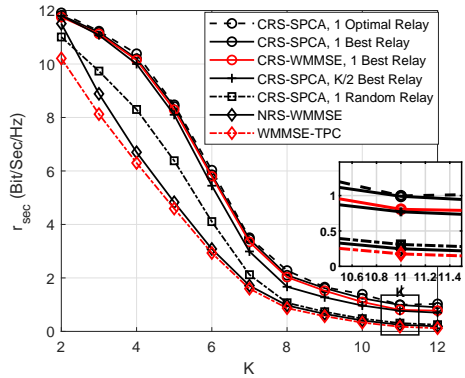}
\par\end{centering}
\caption{The $r_{sec}$ versus $K$ comparison of different strategies, with
$P_{t}=40\textrm{dBm}$ and $N_{t}=4$.}
\end{figure}
Fig. 3 depicts the average convergence of the proposed FIPSA for CRS-SPCA,
CRS-WMMSE, WMMSE-TPC, and NRS-SPCA algorithms using the OF value of
\eqref{eq:38-1}, $\vartheta$ versus the number of iterations. As
observed, the average convergence speed of all algorithm are fast
and converge within a few iterations. Though the proposed FIPSA algorithm
is able to converge quickly within a few iterations. This is similar
to the convergence speed of WMMSE.

Fig. 4 and 5 show the WCSR, $r_{sec}$ achieved by the different strategies
versus the total power budget $P_{t}$ for $N_{t}=4$ and $8$ UAV-BS
transmit antennas. As expected $r_{sec}$ is increased with $P_{t}$.
We can also observe in Fig. 4 that given a total power becomes high,
the $r_{sec}$ is increased. We can also observe that given a specific
total power budget $P_{t}$, BR has almost the same performance for
the proposed CRS-SPCA algorithm as OR. A comparison shows that $\frac{K}{2}$BR,
which is the same as 2BR when $K=4$, does have a negative effect
on WCSR and has a similar  performance to the baseline 1RR . To enhance
the WCSR among users, reducing the size of $\mathcal{K}_{1}$ is preferred,
because then more users can benefit from the cooperative transmission.
Under the given 1BR scheduling protocol, we compare the algorithms
of optimizing the RSMA TPCs, message split variables, time slot sharing
and power allocation. Fig. 4 shows that the proposed CRS-SPCA algorithm
performs similarly or even better in terms of WCSR than all the existing
transmission schemes. 

According to Fig. 5, for $N_{t}=8$ the relative WCSR gain of the
proposed CRS-SPCA increases, because CRS provides improved interference
management capabilities, when the multi-user interference is strong. 

Additionally, we see that when the number of transmit antennas $N_{t}$
gets decreased, the WCSR difference between the suggested CRS-SPCA
(with 1BR) and WMMSE-TPC increases, which it implies that TPC-based
beamformer is only appropriate for under-loaded conditions. As $N_{t}$
decreases, it is unable to handle the multi-user interference caused
by all other users. In comparison, as RS-aided transmission approaches
can partially decode the interference and partially treat interference
as noise, they are more tolerant to the network load. A higher disparity
in channel strength results in a more severe loss of common rate.
By re-transmitting the $s_{c}$ to the worst-case user in CRS, improves
$\mathcal{C}_{k}^{c}$ considerably. Therefore, $\theta$ is substantially
closer to $1$ when user channel strength disparities are negligible.
As a result, the advantages of using CRS over NRS decrease.
\begin{figure}[tbh]
\begin{centering}
\includegraphics[scale=0.64]{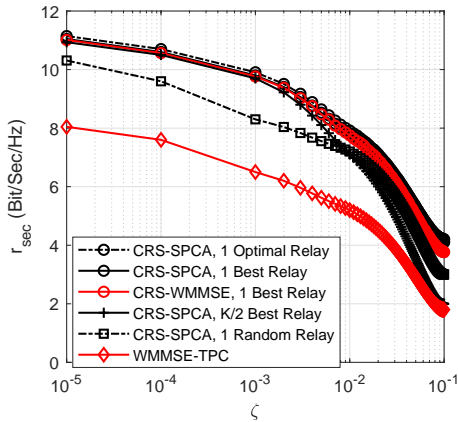}
\par\end{centering}
\caption{The $r_{sec}$ versus $\zeta$ comparison of different strategies,
with $P_{t}=40\textrm{dBm}$ and $N_{t}=4$.}
\end{figure}

\begin{spacing}{0.98}
\noindent Fig. 6 depicts the WCSR $r_{sec}$ versus the number of
users $K$. Since $\stackrel[k=1]{K}{\sum}\left(\begin{array}{c}
K\\
k
\end{array}\right)$ scheduling groups must be considered for obtaining the optimal SRUS,
the complexity escalates as $K$ increases. We observe that in the
$K$-user CRS-assisted transmission network using a single SRUS-based
protocol would suffice.

\noindent As shown in Fig. 6, the gap rate between NRS and CRS clearly
increases as the number of users increases. It follows that as the
number of users grows, the ideal theta decreases. This is due to the
fact that the multi-user interference increases with the number of
users, and a larger portion of the user messages $\left.\mathcal{W}_{k}\right|_{k=1}^{K}$
is transmitted via the common stream $s_{c}$. On the other hand,
then $Eve$ is more likely to eavesdrop successfully on the common
stream which leads to a lower WCSR.
\end{spacing}

To illustrate the robustness of our proposed framework against imperfect
E-CSIT, we have produced Fig. 7, where the average WCSR is depicted
versus the E-CSIT estimation error $\zeta$. Observe that regardless
of the value of $\zeta$, comparing the WCSR of the proposed CRS-SPCA
scheme to that of WMMSE-TPC demostrates the superiority of the CRS-aided
transmission scheme over traditional linear TPC-based beamformer.
Additionally, the WCSR performances of the various scheduling relaying
protocols are similar to each other for low E-CSIT estimation error
of $\zeta$, but at high values of $\zeta$, \textquotedblleft $\frac{K}{2}$BR\textquotedblright{}
performs about $50\%$ worse than \textquotedbl 1BR.\textquotedbl{} 

Finally, Fig. 8 shows the WCSR achieved by the proposed framework
versus the $\theta$ using the \textquotedblleft $1$BR\textquotedblright{}
method. In this experiment we aim for observing the impact of the
UAV-BS altitude $H$ and $\theta$ on the achievable WCSR $r_{sec}$.
It is interesting to note that the $r_{sec}$ vs. $\theta$ curve
is concave, and there is an optimum $\theta^{*}$ at which the $r_{sec}$
will be maximized. This figure also indicates that the WCSR critically
depends on the altitude of the UAV-BS. As the height of the UAV-BS
increases, the value of $\theta$ becomes closer to $1$. This is
because, owing to the LoS links, the quality of the common signal
received at CEU is good enough at the optimal altitude $H^{*}$. Hence
no cooperation is needed for relaying the common stream. In Fig. 8,
increasing the height $H$ and moving away from the optimal altitude
$H^{*}$ results in approximately similar distances between 
\begin{figure}[tbh]
\begin{centering}
\includegraphics[scale=0.7]{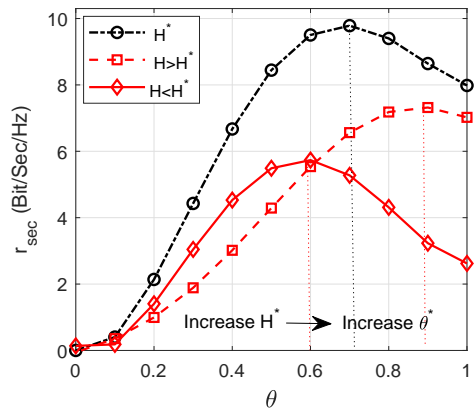}
\par\end{centering}
\caption{The WCSR versus $\theta$ comparison of different $H$, with $P_{t}=40\textrm{dBm}$
and $N_{t}=4$.}
\end{figure}
the UAV-BS and all users, hence the WCSR is reduced due to its higher
path-loss component. By contrast, when the height of the UAV-BS is
reduced from its optimal value, $\theta^{*}$ decreases as well due
to the emergence of fading. As a result, the CEU's common signal is
adversely affected, and reliance on the cooperative phase becomes
essential. 

\section{Conclusion\label{sec:VI}}

To conclude, we studied the robust and secure max-min fairness of
cooperative multi-user RSMA in a MISO-BC UAV network downlink where
only imperfect E-CSIT is available. We formulated the problem of maximizing
the WCSR by jointly optimizing the SRUS and the network's resources
allocation, including the TPCs, time slot sharing and power allocation.
To circumvent the non-convexity resulting from the discrete nature
of th SRUS, we proposed a two-stage algorithm, where the SRUS and
network resources optimization were performed in two consecutive stages.
As for the SRUS, we analytically showed that we only need the ratio
of the $\textrm{UAV-BS}\rightarrow\left\{ \left.U_{k}\right|_{k\in\mathcal{K}}\right\} $
and $\left\{ \left.U_{k}\right|_{k\in\mathcal{K}}\right\} \rightarrow\left\{ Eve\right\} $
channel gains for two type of centralized and distributed protocols.
On the other hand, an SPCA-based solution has been proposed to cope
with the resultant non-convexity imposed by the network resource allocation.
Our numerical results show that by applying the proposed solution,
the WCSR is substantially boosted over that of the benchmarks. Thus,
we conclude that our cooperative multi-user RSMA framework is capable
of improving the confidentiality of 6G networks. 

\appendices{}

\section{Proof of Proposition \ref{prop:At-the-global}}

We first prove the equivalence of the common secrecy rate achieved
by users in $\mathcal{K}_{1}$ and $\mathcal{K}_{2}$ at the globally
optimal point \eqref{eq:19} by the method of contradiction. By assuming
that $k_{1,min}$ and $k_{2,min}$ are the two users that respectively
achieve the worst common secrecy rate in $\mathcal{K}_{1}$ and $\mathcal{K}_{2}$
at the globally optimal point $\left(\mathbf{P}^{*},\theta^{*},\mathbf{\mathbf{\boldsymbol{\chi}}}^{*},\mathcal{K}_{1}^{*},\left\{ p_{j}^{*}\right\} _{\forall j\in\mathcal{K}_{1}^{*}},p_{z}^{*}\right)$,
we obtain \eqref{eq:49}, where 
\begin{figure*}[tbh]
\begin{flalign}
\mathcal{R}_{c,\mathcal{K}_{1}}^{sec^{*}} & \triangleq\theta^{*}\left[\gamma_{c,k_{1,min}}^{(1)}\left(\mathbf{P}^{*}\right)-\gamma_{c,e}^{(1)}\right]^{+}=\theta^{*}\gamma_{c,k_{1,min}}^{\text{sec}(1)}\left(\mathbf{P}^{*}\right),\label{eq:49}\\
\mathcal{R}_{c,\mathcal{K}_{2}}^{sec^{*}} & \triangleq\theta^{*}\left[\gamma_{c,k_{2,min}}^{(1)}\left(\mathbf{P}^{*}\right)-\gamma_{c,e}^{(1)}\right]^{+}+\left(1-\theta^{*}\right)\left[\gamma_{c,k_{2,min}}^{(2)}\left(\left\{ p_{j}^{*}\right\} _{\forall j\in\mathcal{K}_{1}^{*}},p_{z}^{*},\mathcal{K}_{1}^{*}\right)-\gamma_{c,\text{e}}^{(2)}\right]^{+}\nonumber \\
\, & =\theta^{*}\gamma_{c,k_{2,min}}^{\text{sec}(1)}\left(\mathbf{P}^{*}\right)+\left(1-\theta^{*}\right)\gamma_{c,k_{2,min}}^{\text{sec}(2)}\left(\left\{ p_{j}^{*}\right\} _{\forall j\in\mathcal{K}_{1}^{*}},p_{z}^{*},\mathcal{K}_{1}^{*}\right),\nonumber \\
\, & =\theta^{*}\left(\gamma_{c,k_{2,min}}^{\text{sec}(1)}\left(\mathbf{P}^{*}\right)-\gamma_{c,k_{2,min}}^{\text{sec}(2)}\left(\left\{ p_{j}^{*}\right\} _{\forall j\in\mathcal{K}_{1}^{*}},p_{z}^{*},\mathcal{K}_{1}^{*}\right)\right)+\gamma_{c,k_{2,min}}^{\text{sec}(2)}\left(\left\{ p_{j}^{*}\right\} _{\forall j\in\mathcal{K}_{1}^{*}},p_{z}^{*},\mathcal{K}_{1}^{*}\right),\nonumber 
\end{flalign}

\begin{tabular}{c}
~~~~~~~~~~~~~~~~~~~~~~~~~~~~~~~~~~~~~~~~~~~~~~~~~~~~~~~~~~~~~~~~~~~~~~~~~~~~~~~~~~~~~~~~~~~~~~~~~~~~~~~~~~~~~~~~~~~~~~~~~~~~~~~~~~~~~~~~~~~~~~~~~~~\tabularnewline
\hline 
\end{tabular}
\end{figure*}
\begin{equation}
\gamma_{c,n}^{(1)}\left(\mathbf{P}^{*}\right)\triangleq\log_{2}\left(1+\frac{\left|{\mathbf{h}_{n}}^{H}\mathbf{p}_{c}^{*}\right|^{2}}{\underset{k\in\mathcal{K}}{\sum}\left|{\mathbf{h}_{n}}^{H}\mathbf{p}_{k}^{*}\right|^{2}+\sigma_{n}^{2}}\right),\label{eq:50}
\end{equation}
\begin{equation}
\gamma_{c,n}^{(2)}\left(\left\{ p_{j}^{*}\right\} _{\forall j\in\mathcal{K}_{1}^{*}},p_{z}^{*},\mathcal{K}_{1}^{*}\right)\triangleq\log_{2}\left(1+\frac{\underset{j\in\mathcal{K}_{1}^{*}}{\sum}p_{j}^{*}\left|{h_{j,n}}\right|^{2}}{p_{z}^{*}\left|{\mathbf{h}_{n}}^{H}\mathbf{\hat{p}}_{z}\right|^{2}+\sigma_{n}^{2}}\right)\label{eq:51}
\end{equation}

\begin{figure*}[tbh]
\begin{equation}
\theta^{*}\gamma_{c,k_{1,min}}^{\text{sec}(1)}\left(\mathbf{P}^{*}\right)>\theta^{*}\left(\gamma_{c,k_{2,min}}^{\text{sec}(1)}\left(\mathbf{P}^{*}\right)-\gamma_{c,k_{2,min}}^{\text{sec}(2)}\left(\left\{ p_{j}^{*}\right\} _{\forall j\in\mathcal{K}_{1}^{*}},p_{z}^{*},\mathcal{K}_{1}^{*}\right)\right)+\gamma_{c,k_{2,min}}^{\text{sec}(2)}\left(\left\{ p_{j}^{*}\right\} _{\forall j\in\mathcal{K}_{1}^{*}},p_{z}^{*},\mathcal{K}_{1}^{*}\right)\label{eq:52}
\end{equation}
\begin{equation}
\Longleftrightarrow\theta^{*}>\frac{\gamma_{c,k_{2,min}}^{\text{sec}(2)}\left(\left\{ p_{j}^{*}\right\} _{\forall j\in\mathcal{K}_{1}^{*}},p_{z}^{*},\mathcal{K}_{1}^{*}\right)}{\gamma_{c,k_{1,min}}^{\text{sec}(1)}\left(\mathbf{P}^{*}\right)-\gamma_{c,k_{2,min}}^{\text{sec}(1)}\left(\mathbf{P}^{*}\right)+\gamma_{c,k_{2,min}}^{\text{sec}(2)}\left(\left\{ p_{j}^{*}\right\} _{\forall j\in\mathcal{K}_{1}^{*}},p_{z}^{*},\mathcal{K}_{1}^{*}\right)}=\bar{\theta},\label{eq:53}
\end{equation}
\begin{flalign}
\theta^{*}\left[\gamma_{c,k_{1,min}}^{(1)}\left(\mathbf{P}^{*}\right)-\gamma_{c,e}^{(1)}\right]^{+}=\theta^{*}\left[\gamma_{c,k_{2,min}}^{(1)}\left(\mathbf{P}^{*}\right)-\gamma_{c,e}^{(1)}\right]^{+}+\left(1-\theta^{*}\right)\left[\gamma_{c,k_{2,min}}^{(2)}\left(\left\{ p_{j}^{*}\right\} _{\forall j\in\mathcal{K}_{1}^{*}},p_{z}^{*},\mathcal{K}_{1}^{*}\right)-\gamma_{c,\text{e}}^{(2)}\right]^{+},\label{eq:54}
\end{flalign}
\begin{flalign}
\theta^{*}\left[\gamma_{c,k_{1,min}}^{(1)}-\gamma_{c,k_{2,min}}^{(1)}\right]=\left(1-\theta^{*}\right)\left[\gamma_{c,k_{2,min}}^{(2)}-\gamma_{c,\text{e}}^{(2)}\right]^{+},\,\,\, & R_{c,\mathcal{K}_{1}}^{*}-R_{c,\mathcal{K}_{2}}^{\text{*}(1)}=\mathcal{R}_{c,\mathcal{K}_{2}}^{\text{sec}*},\label{eq:55}
\end{flalign}

\begin{tabular}{c}
~~~~~~~~~~~~~~~~~~~~~~~~~~~~~~~~~~~~~~~~~~~~~~~~~~~~~~~~~~~~~~~~~~~~~~~~~~~~~~~~~~~~~~~~~~~~~~~~~~~~~~~~~~~~~~~~~~~~~~~~~~~~~~~~~~~~~~~~~~~~~~~~~~~\tabularnewline
\hline 
\end{tabular}
\end{figure*}
Note that for the weakest legitimate user in $\mathcal{K}_{2}$, i.e.,
$U_{k_{2,min}}$, we must have $\gamma_{c,k_{2,min}}^{\text{sec}(1)}\left(\mathbf{P}^{*}\right)<\gamma_{c,k_{2,min}}^{\text{sec}(2)}\left(\left\{ p_{j}^{*}\right\} _{\forall j\in\mathcal{K}_{1}^{*}},p_{z}^{*},\mathcal{K}_{1}^{*}\right)$
when $0<\theta^{*}<1$. Otherwise, both $\mathcal{R}_{c,\mathcal{K}_{1}}^{\text{sec}^{*}}$
and $\mathcal{R}_{c,\mathcal{K}_{2}}^{\text{sec}^{*}}$ are increasing
functions of $\theta$ for the optimal $\mathbf{P}^{*},\left\{ p_{j}^{*}\right\} _{\forall j\in\mathcal{K}_{1}^{*}},p_{z}^{*},\mathcal{K}_{1}^{*}$,
and the optimal $\theta^{*}$ should be $1$ in order to maximize
the minimum secrecy rate among users (\textbf{\textcolor{black}{Note}}:
$\theta^{*}<1$ means that there are some users, i.e., $\mathcal{K}_{2}^{*}\neq\phi$,
that can not receive the common message with good quality and hence
have to receive it in the cooperative phase. Hence, when $0<\theta<1$,
$\mathcal{R}_{c,\mathcal{K}_{1}}^{\text{sec}^{*}}$ is a monotonically
increasing function of $\theta$, while $\mathcal{R}_{c,\mathcal{K}_{2}}^{\text{sec}^{*}}$
is a monotonic decreasing function of $\theta$.
\begin{IEEEproof}
If at the $\theta^{*}$, we assume $\mathcal{R}_{c,\mathcal{K}_{1}}^{\text{sec}^{*}}>$$\mathcal{R}_{c,\mathcal{K}_{2}}^{\text{sec}^{*}}$,
then we have \eqref{eq:52} and \eqref{eq:53}.

By decreasing $\theta$ from $\theta^{*}$ to $\bar{\theta}$, we
have $\mathcal{R}_{c,2}^{\text{sec}*}\left(\bar{\theta}\right)$$>$$\mathcal{R}_{c,2}^{\text{sec}*}\left(\theta^{*}\right)$
and $\mathcal{R}_{c,1}^{\text{sec}*}\left(\bar{\theta}\right)$$=$
$\mathcal{R}_{c,2}^{\text{sec}*}\left(\bar{\theta}\right)$. Then
the achievable common secrecy rate of $\underset{\forall k\in\mathcal{K}}{\sum}$
$\bar{C}_{k}^{\text{sec}}=$ $\min\left\{ {\mathcal{R}_{c,1}^{\text{sec}^{*}}\left(\bar{\theta}\right)},\right.\left.{\mathcal{R}_{c,2}^{\text{sec}^{*}}\left(\bar{\theta}\right)}\right\} $
increases. By allocating the improved common secrecy rate to the worst-case
user, the value of the objective function achieved by using the new
solution $\left(\mathbf{P}^{*},\bar{\theta},\bar{\mathbf{\mathbf{\boldsymbol{\chi}}}},\mathcal{K}_{1}^{*},\left\{ p_{j}^{*}\right\} _{\forall j\in\mathcal{K}_{1}^{*}},p_{z}^{*}\right)$
is higher than that of $\left(\mathbf{P}^{*},\theta^{*},\mathbf{\mathbf{\boldsymbol{\chi}}}^{*},\mathcal{K}_{1}^{*},\left\{ p_{j}^{*}\right\} _{\forall j\in\mathcal{K}_{1}^{*}},p_{z}^{*}\right)$,
which contradicts to the fact that $\left(\mathbf{P}^{*},\theta^{*},\mathbf{\mathbf{\boldsymbol{\chi}}}^{*},\mathcal{K}_{1}^{*},\left\{ p_{j}^{*}\right\} _{\forall j\in\mathcal{K}_{1}^{*}},p_{z}^{*}\right)$
is the globally optimal point. Similarly, we obtain that if $\mathcal{R}_{c,\mathcal{K}_{1}}^{\text{sec}^{*}}$
$<$ $\mathcal{R}_{c,\mathcal{K}_{2}}^{\text{sec}^{*}}$, $\theta^{*}<\bar{\theta}$
holds. A better solution is obtained by decreasing $\theta$ from
$\theta^{*}$ to $\bar{\theta}$, and then a contradiction arises.
Hence, we draw the conclusion that at the globally optimal point $\left(\mathbf{P}^{*},\theta^{*},\mathbf{\mathbf{\boldsymbol{\chi}}}^{*},\mathcal{K}_{1}^{*},\left\{ p_{j}^{*}\right\} _{\forall j\in\mathcal{K}_{1}^{*}},p_{z}^{*}\right)$
of problem \eqref{eq:18}, $\mathcal{R}_{c,\mathcal{K}_{1}}^{\text{sec}^{*}}$=$\mathcal{R}_{c,\mathcal{K}_{2}}^{\text{sec}^{*}}$.
As $\underset{k\in\mathcal{K}_{2}^{*}}{\min}$$\left\{ \mathcal{R}_{c,k}^{\text{sec}(1)}+\right.\left.\mathcal{R}_{c,k}^{\text{sec}(2)}\right\} $
$\geq$ $\underset{k\in\mathcal{K}_{2}^{*}}{\min}\left\{ \mathcal{R}_{c,k}^{\text{sec}(1)}\right\} $
$+$ $\underset{k\in\mathcal{K}_{2}^{*}}{\min}$ $\left\{ \mathcal{R}_{c,k}^{\text{sec}(2)}\right\} $,
we obtain that $\mathcal{R}_{c,\mathcal{K}_{2}}^{\text{sec}^{*}}$
$>$$\underset{k\in\mathcal{K}_{2}^{*}}{\min}$ $\left\{ \mathcal{R}_{c,k}^{\text{sec}(1)}\right\} $.
Based on \eqref{eq:19}, we have $\mathcal{R}_{c,\mathcal{K}_{1}}^{\text{sec}^{*}}>$
$\underset{k\in\mathcal{K}_{2}^{*}}{\min}\left\{ \mathcal{R}_{c,k}^{\text{sec}(1)}\right\} $
when $0<\theta^{*}<1$ and the proof of proposition $1$ is completed.
\end{IEEEproof}

\section{\label{sec:Proof-of-Proposition-conv}Proof of Proposition \ref{prop:For-any-feasible}}

SPCA ensures monotonic improvement of $r_{sec}$, i.e., $r_{sec}^{[n]}\geq r_{sec}^{[n+1]}$.
This is due to the fact that the solution generated by solving problem
(\ref{eq:22}) at iteration $\left[n-1\right]$ is a feasible point
of problem (\ref{eq:22}) at iteration $\left[n\right]$. Due to the
transmit power constraint (\eqref{eq:18}-$C_{4}$), the sequence
$\left\{ r_{sec}^{[n]}\right\} _{n=1}^{\infty}$ is bounded from above,
which implies that the convergence of the proposed SPCA-based algorithm
is guaranteed. Next, we show that the sequence of $\left(\mathbf{P}^{[n]},\theta^{[n]},\mathbf{\mathbf{\boldsymbol{\chi}}}^{[n]},\left\{ p_{j}^{[n]}\right\} _{\forall j\in\mathcal{K}_{1}^{*}},\boldsymbol{\mathbf{\beta}}_{\left\{ c,e,p\right\} }^{[n]},\boldsymbol{\mathbf{\mathbf{\rho}}}_{\left\{ c,p\right\} }^{[n]},\alpha_{c,e}^{[n]}\right)$
converges to the set of stationary points of problem (\ref{eq:22}).
The proposed SPCA-based algorithm is in fact an inner approximation
algorithm of  the non-convex optimization literature \cite{28,29}.
This is proved by showing the equivalence of the KKT conditions of
problem (\ref{eq:21}) and problem (\ref{eq:22}) when the solution
$\left(\mathbf{P},\theta,\mathbf{\mathbf{\boldsymbol{\chi}}},\left\{ p_{j}\right\} _{\forall j\in\mathcal{K}_{1}^{*}},\boldsymbol{\mathbf{\beta}}_{\left\{ c,e,p\right\} },\boldsymbol{\mathbf{\mathbf{\rho}}}_{\left\{ c,p\right\} },\alpha_{c,e}\right)$
is equal to $\left(\mathbf{P}^{[n]},\theta^{[n]},\mathbf{\mathbf{\boldsymbol{\chi}}}^{[n]},\left\{ p_{j}^{[n]}\right\} _{\forall j\in\mathcal{K}_{1}^{*}},\boldsymbol{\mathbf{\beta}}_{\left\{ c,e,p\right\} }^{[n]},\boldsymbol{\mathbf{\mathbf{\rho}}}_{\left\{ c,p\right\} }^{[n]},\alpha_{c,e}^{[n]}\right)$.
Combined with the fact that the Taylor approximations made in (\ref{eq:22})
are asymptotically tight as $n\rightarrow\infty$ \cite{29}, we can
see that the solution of the proposed SPCA-based algorithm converges
to the set of KKT points (which are also known as the stationary points)
of problem (\ref{eq:21}).

\end{document}